\documentstyle[11pt]{article}     

\expandafter\ifx\csname amssym.def\endcsname\relax \else\endinput\fi
\expandafter\edef\csname amssym.def\endcsname{%
       \catcode`\noexpand\@=\the\catcode`\@\space}
\catcode`\@=11
\def\undefine#1{\let#1\undefined}
\def\newsymbol#1#2#3#4#5{\let\next@\relax
 \ifnum#2=\@ne\let\next@\msafam@\else
 \ifnum#2=\tw@\let\next@\msbfam@\fi\fi
 \mathchardef#1="#3\next@#4#5}
\def\mathhexbox@#1#2#3{\relax
 \ifmmode\mathpalette{}{\m@th\mathchar"#1#2#3}%
 \else\leavevmode\hbox{$\m@th\mathchar"#1#2#3$}\fi}
\def\hexnumber@#1{\ifcase#1 0\or 1\or 2\or 3\or 4\or 5\or 6\or 7\or 8\or
 9\or A\or B\or C\or D\or E\or F\fi}

\font\tenmsa=msam10
\font\sevenmsa=msam7
\font\fivemsa=msam5
\newfam\msafam
\textfont\msafam=\tenmsa
\scriptfont\msafam=\sevenmsa
\scriptscriptfont\msafam=\fivemsa
\edef\msafam@{\hexnumber@\msafam}
\mathchardef\dabar@"0\msafam@39
\def\dashrightarrow{\mathrel{\dabar@\dabar@\mathchar"0\msafam@4B}}
\def\dashleftarrow{\mathrel{\mathchar"0\msafam@4C\dabar@\dabar@}}

\def\ulcorner{\delimiter"4\msafam@70\msafam@70 }
\def\urcorner{\delimiter"5\msafam@71\msafam@71 }
\def\llcorner{\delimiter"4\msafam@78\msafam@78 }
\def\lrcorner{\delimiter"5\msafam@79\msafam@79 }
\def\yen{{\mathhexbox@\msafam@55}}
\def\checkmark{{\mathhexbox@\msafam@58}}
\def\circledR{{\mathhexbox@\msafam@72}}
\def\maltese{{\mathhexbox@\msafam@7A}}

\font\tenmsb=msbm10
\font\sevenmsb=msbm7
\font\fivemsb=msbm5
\newfam\msbfam
\textfont\msbfam=\tenmsb
\scriptfont\msbfam=\sevenmsb
\scriptscriptfont\msbfam=\fivemsb
\edef\msbfam@{\hexnumber@\msbfam}
\def\Bbb#1{{\fam\msbfam\relax#1}}
\def\widehat#1{\setbox\z@\hbox{$\m@th#1$}%
 \ifdim\wd\z@>\tw@ em\mathaccent"0\msbfam@5B{#1}%
 \else\mathaccent"0362{#1}\fi}
\def\widetilde#1{\setbox\z@\hbox{$\m@th#1$}%
 \ifdim\wd\z@>\tw@ em\mathaccent"0\msbfam@5D{#1}%
 \else\mathaccent"0365{#1}\fi}
\font\teneufm=eufm10
\font\seveneufm=eufm7
\font\fiveeufm=eufm5
\newfam\eufmfam
\textfont\eufmfam=\teneufm
\scriptfont\eufmfam=\seveneufm
\scriptscriptfont\eufmfam=\fiveeufm
\def\frak#1{{\fam\eufmfam\relax#1}}

\csname amssym.def\endcsname

\makeatletter
\def\section{\@startsection {section}{1}{\z@}{-3.5ex plus -1ex minus 
 -.2ex}{2.3ex plus .2ex}{\large\sc}}
\def\subsection{\@startsection{subsection}{2}{\z@}{-3.25ex plus -1ex minus 
 -.2ex}{1.5ex plus .2ex}{\normalsize\sc}}
\makeatother

\makeatletter
\@addtoreset{equation}{section}                        

\makeatother

\newcommand{\nc}{\newcommand}
\newcommand{\rnc}{\renewcommand}

\nc{\be}{\begin{equation}}
\nc{\ee}{\end{equation}}
\nc{\bea}{\begin{eqnarray}}
\nc{\eea}{\end{eqnarray}}

\nc{\trac}[2]{{\textstyle\frac{#1}{#2}}}

\nc{\ex}[1]{\mbox{e}^{\,\textstyle#1}}

\nc{\CC}{\Bbb{C}}
\nc{\PP}{\Bbb{P}}
\nc{\RR}{\Bbb{R}}
\nc{\ZZ}{\Bbb{Z}}
\nc{\II}{\Bbb{I}}

\rnc{\a}{\alpha}
\nc{\al}{\a^{l}}
\rnc{\d}{\delta}
\nc{\ga}{\gamma}
\nc{\la}{\lambda}
\nc{\lal}{\la_{l}}
\nc{\f}{\phi}

\nc{\ab}{\bar{\a}}
\nc{\ap}{\a^{+}}
\nc{\abm}{\ab^{-}}
\rnc{\b}{\beta}
\nc{\bb}{\bar{\b}}
\nc{\bbp}{\bb_{z}^{+}}
\nc{\bm}{\b_{z}^{-}}
\rnc{\gg}{\gamma}
\nc{\fb}{\bar{\phi}}
\nc{\vf}{\varphi}
\nc{\p}{\psi}
\def\pb{\bar{\psi}}
\rnc{\c}{\chi}
\nc{\m}{\mu}
\nc{\n}{\nu}
\rnc{\o}{\omega}
\nc{\Om}{\Omega}
\rnc{\t}{\theta}
\nc{\eps}{\epsilon}
\nc{\F}{\Phi}

\nc{\ad}{\mathop{\mbox{ad}}\nolimits}
\nc{\tr}{\mathop{\mbox{tr}}\nolimits}
\nc{\Tr}{\mathop{\mbox{Tr}}\nolimits}
\nc{\Det}{\mathop{\mbox{Det}}\nolimits}
\nc{\rk}{\mathop{\mbox{rk}}\nolimits}
\nc{\diag}{\mbox{diag}}
\nc{\ra}{\rightarrow}
\nc{\Ra}{\Rightarrow}
\nc{\LRa}{\Leftrightarrow}
\nc{\ot}{\otimes}
\rnc{\ss}{\subset}
\nc{\nul}{\noindent\underline}
\nc{\non}{\nonumber\\}
\rnc{\S}{\Sigma}
\nc{\tp}{2\pi i}
\nc{\del}{\partial}
\nc{\dbar}{\bar{\del}}

\nc{\sun}{SU(n)}
\nc{\sumn}{SU(m+n)}
\nc{\sumun}{S(U(m)\times U(n))}
\nc{\cpm}{\CC\PP(m)}
\nc{\cpn}{\CC\PP(n)}
\nc{\gmn}{G(m,m+n)}
\nc{\zm}{z_{m}}
\nc{\zn}{z_{n}}
\nc{\zmn}{z_{m+n}}
\nc{\Zm}{\ZZ_{m}}
\nc{\Zn}{\ZZ_{n}}
\nc{\Zmn}{\ZZ_{m+n}}
\rnc{\Im}{\II_{m}}
\nc{\In}{\II_{n}}
\nc{\mn}{(m|n)}
\nc{\hm}{h_{1}^{(m)}}
\nc{\hn}{h_{2}^{(n)}}
\nc{\fm}{\f^{(m)}}
\nc{\fn}{\f^{(n)}}
\nc{\grm}{\Gamma^{r,(m)}}
\nc{\grn}{\Gamma^{r,(n)}}
\nc{\gwm}{\Gamma^{w,(m)}}
\nc{\gwn}{\Gamma^{w,(n)}}
\nc{\kmn}{k+m+n}
\nc{\lam}{\la^{(m)}}
\nc{\lan}{\la^{(n)}}
\nc{\am}{\a^{(m)}}
\nc{\an}{\a^{(n)}}

\rnc{\lg}{\frak{g}}
\nc{\lt}{\frak{t}}
\nc{\lk}{\frak{k}}
\nc{\lh}{\frak{h}}
\nc{\pik}{\Pi_{\lk}}
\nc{\pip}{\Pi_{+}}
\nc{\pim}{\Pi_{-}}
\nc{\pih}{\Pi_{\lh}}
\nc{\jz}{J_{z}}
\nc{\jzh}{\jz^{\lh}}
\nc{\jzp}{\jz^{+}}
\nc{\jzm}{\jz^{-}}
\nc{\dz}{\del_{z}}
\nc{\dzb}{\del_{\bar{z}}}
\nc{\zb}{\bar{z}}
\nc{\az}{A_{z}}
\nc{\azb}{A_{\bar{z}}}
\nc{\g}{g^{-1}}
\nc{\dw}{\Delta_{W}}
\nc{\Ad}{{\mbox{Ad}}}
\nc{\ks}{Ka\-za\-ma-\-Su\-zu\-ki}
\nc{\KS}{\ks}
\nc{\ksm}{\ks\ model}
\nc{\cp}[1]{\CC\PP(#1)}
\nc{\gmnk}{\gmn_{k}}
\nc{\cO}{{\cal O}}
\nc{\bcO}{\bar{\cO}}
\nc{\bO}{\bar{O}}

\nc{\subs}[1]{{\vspace*{0.5cm}}%
{\noindent\underline{#1}}{\addcontentsline{toc}{subsection}{#1}}%
{\vspace*{0.3cm}}}

\begin{document}
\global\parskip=4pt

\makeatletter
\begin{titlepage}
\begin{center}
{\LARGE\bf Grassmannian Topological Kazama-Suzuki\\[4mm] 
Models and Cohomology}\\
\vskip .3in
{\bf Matthias Blau}\footnote{e-mail: mblau@enslapp.ens-lyon.fr; supported 
by EC Human Capital and Mobility Grant ERB-CHB-GCT-93-0252}
\vskip .10in
Laboratoire de Physique Th\'eorique
{\sc enslapp}\footnote{URA 14-36 du CNRS,
associ\'ee \`a l'E.N.S. de Lyon,
et \`a l'Universit\'e de Savoie}\\
ENSLyon,
46 All\'ee d'Italie,\\
{}F-69364 Lyon CEDEX 07, France\\
\vskip .20in
{\bf Faheem Hussain}\footnote{e-mail: hussainf@ictp.trieste.it}  and 
{\bf George Thompson}\footnote{e-mail: thompson@ictp.trieste.it}\\
\vskip .10in
ICTP\\
P.O. Box 586\\
34014 Trieste, Italy
\end{center}
\begin{small}
\noindent
We investigate in detail the topological gauged Wess-Zumino-Witten models
describing topological Kazama-Suzuki models based on complex Grassmannians. 
We show that there is a topological sector in which the ring of 
observables (constructed from the Grassmann odd scalars of the theory) 
coincides with the classical cohomology ring of the Grassmannian
for all values of the level $k$. We perform a detailed analysis of the
non-trivial topological sectors arising from the adjoint gauging,
and investigate the general ring 
structure of bosonic correlation functions, uncovering a whole hierarchy
of level-rank relations (including the standard level-rank duality)
among models based on different Grassmannians. Using the previously established
localization of the topological Kazama-Suzuki model to an Abelian topological
field theory, we reduce the correlators to finite-dimensional purely algebraic
expressions. As an application, 
these are evaluated explicitly for the $\cp{2}$ model at level $k$ and
shown for all $k$ to coincide with the cohomological intersection numbers 
of the two-plane Grassmannian $G(2,k+2)$, thus realizing the level-rank
duality between this model and the $G(2,k+2)$ model at level one.
\end{small}
\vfill
IC/95/341 \hfill hep-th/9510194 \hfill
{\small E}N{\large S}{\Large L}{\large A}P{\small P}-L-557/95
\end{titlepage}
\makeatother

\begin{small}
\tableofcontents
\end{small}

\setcounter{footnote}{0}

\section{Introduction}

In \cite{ks1}, Kazama and Suzuki 
discovered a large new class of $N=2$ superconformal
field theories (SCFTs) by investigating under which conditions
the $N=1$ super-GKO construction \cite{gko} for a $G/H$ coset 
conformal field theory actually gives rise to an $N=2$ SCFT. Some
aspects of these \ks\ models, among which are the particularly interesting 
cosets with $\mbox{rk}\;G = \mbox{rk}\;H$ and $G/H$ K\"ahler, have been
investigated both in the context of $N=2$ SCFT \cite{lvw,flmw,lw,dg1,dg2,ht}
and in the context of string model building \cite{string}. These
investigations have unravelled the rich and beautiful structure of
the \ks\ models arising from the interplay between the geometry and 
topology of coset spaces on the one hand and the $N=2$ superconformal 
algebra on the other. 

Certain global aspects of (two-dimensional) field theories are, however, 
frequently easier to extract from an action-based path integral formulation
of the theory. Such a Lagrangian realization of \ks\ models was provided
in \cite{ewcp,nak}. There it was shown that, under the conditions for
$N=2$ superconformal symmetry determined in \cite{ks1}, also the 
Lagrangian realization of the $N=1$ coset models 
as supersymmetric gauged Wess-Zumino-Witten (WZW) models \cite{schnitzer}
actually possesses the expected $N=2$ symmetry. 

This Lagrangian realization 
was investigated further by Nakatsu and Sugawara \cite{ns} who, in 
particular, showed how with a suitable change of variables a direct link
could be established between the gauge-theoretic and the more standard
conformal field theory realization of these models. The Lagrangian
realization of \ks\ models was also used by Henningson in \cite{hen1,hen2}
to discuss other aspects of \ks\ models which are more transparent in
this framework, namely the elliptic genus and mirror symmetry.

Our starting point for this and the companion paper \cite{bhtks} is
the work of Witten \cite{ewcp} who analyzed in detail the topologically
twisted $G/H =\cp{1}$ models (a.k.a.\ $N=2$ minimal models) and their
coupling to topological gravity. In \cite{bhtks}, a large part of
Witten's analysis was generalized to arbitrary topological $G/H$ \ks\ models.

One of the new features of these models is the presence of
topologically non-trivial sectors emerging from the fact that the gauge
group of the model is not $H$ but rather, because of the adjoint
action, $H/Z$ where $Z=Z(G)\cap H$. It was pointed out by Hori
\cite{hori} (in the case of bosonic coset models) that taking into
account the twisted sectors should be tantamount to a resolution of
what is known as the `fixed point problem' \cite{fixedpoint} in the
conformal field theory context. Thus, in \cite{bhtks}, building on work
by Gawedzki \cite{gawedzki} and Hori \cite{hori}, we presented a
generalization of some of the standard properties and constructions of
coset models (Polyakov-Wiegmann identities, wave functions, holomorphic
factorization, \ldots) to the topologically non-trivial situations one
encounters in \ks\ models. Using these results, it was then possible to
show that the techniques of localization \cite{ewcp} and
diagonalization \cite{btver} are still applicable in the present
context.

In particular, it was shown that the path integral of the topological
$G/H$ model can be localized to that of a bosonic $H/H$ model (plus
quantum corrections coming from the chiral anomaly of the $G/H$ model).
More precisely, one should perhaps refer to these as $G/\mbox{Ad}H$ 
or $G/(H/Z)$ and $H/(H/Z)$ models. But in any case, 
by the results of \cite{btver}, the latter can in turn be reduced to
an Abelian topological field theory, namely a $T/T$ model, where $T$ is
the maximal torus of $H$. 

At this stage one has simplified the original
non-linear and non-Abelian supersymmetric model to such an extent that
explicit calculations of correlation functions become not only feasible
but even, in some cases, straightforward. The purpose of the present
paper is to gain a better understanding of the (topological) \ks\
models by performing some explicit calculations in the models based on
complex Grassmannians $G(m,m+n)$.  

It is, more or less, known from the conformal field theory literature
that the chiral $(c,c)$ ring \cite{lvw} of these models at level $k=1$ (which
becomes the observable ring of the topological model) is just the
classical (Dolbeault) cohomology ring \cite{lvw}-\cite{dg2}.

In the present context it is straightforward to establish that there is
indeed a topological sector of the model (the trivial sector) in which
the correlation functions of suitable observables (constructed from
the Grassmann odd scalars of the twisted model) calculate the 
cohomology ring (intersection numbers) of the Grassmannian $G(m,m+n)$.
This can be established without making use of the
simplifying features (localizability, diagonalizability) of the theory
mentioned above. 
Remarkably, with a suitable $k$-dependent normalization of the correlation 
functions, the fact that these correlation functions reproduce the classical
cohomology relations is independent of the level $k$. In this case, we
also find an {\em a priori}
geometrical explanation for the emergence of cohomology
from the \ks\ model by relating the observables to the classical geometry 
of vector bundles on $\gmn$. 

\ks\ models are also known to possess
a level-rank duality symmetry which, in the case at hand, amounts to
the statement that the models one obtains by permuting $k$ and $m$ and $n$
are all related. Analyzing the ring structure in general $\gmn$ level $k$
models, we find here that this is but a particular case of a whole
hierarchy of relations between correlation functions in models based 
on projective spaces and more general Grassmannians, see (\ref{id3}). 
These relations 
take the form of morphisms between subrings of the chiral rings of 
different Grassmannian models.

Now, level-rank duality (see e.g.\ \cite{hslrd,jijnhs,jfcslrd})
is something that is usually not
particularly transparent in the path integral framework. However, 
calculations at higher level (which tend to be difficult in conformal
field theory) are usually not more difficult than low-level calculations
in this framework and thus provide a useful check on level-rank duality. 
In particular, having reduced a general $\gmn$
correlation function to a finite dimensional algebraic expression,
we find (and prove in detail for
$m=2$) that the chiral ring of the $\cpm$ model at level $k$ (in the 
bosonic sector in which there are no fermionic zero modes) is the
cohomology ring of the Grassmannian $G(m,m+k)$, thus realizing the
expected equivalence $G(m,m+1)_{k}\equiv G(m,m+k)_{1}$. That the
path integral approach is, in a sense, `level-rank dual' to the
conformal field theory approach is also epitomized by the fact that
the emergence of the classical cohomology ring directly from the 
level $k=1$ Grassmannian models in the bosonic sector
is slightly less transparent here.

This paper is organized as follows: In section 2 we introduce the
topological \ks\ models and recall the relevant results from
\cite{bhtks}, in particular as regards the localization and
diagonalization of these models. We also make some general comments on
fermionic zero modes and ghost number anomalies (selection rules) for
hermitian symmetric models.

In order to learn how to take into account the topologically non-trivial
sectors, we investigate in some detail the global properties of
the gauge group $H/Z = \sumun/\Zmn$ in the appendix. Based on that,
in section 3 we derive explicit expressions for
the Chern classes of $H/Z$-bundles and the corresponding Dolbeault-index.

In section 4, we work out in detail the general structure of the
(abelianized)
Grassmannian \ks\ models. In section 5 we review the classical description 
of the cohomology of Grassmannians and show that in the topologically trivial
sector (with zero modes of the Grassmann odd scalars) the correlation 
functions reproduce this cohomology. In sections 6 and 7 
we consider bosonic 
correlation functions, show how they can in general be reduced to purely
algebraic finite-dimensional expressions, discuss level-rank dualities
 and work out in detail the equivalence of the correlation functions 
of the $\cp{2}_{k}$ model with the cohomology of $G(2,k+2)$.

There are, of course, a large number of things that still remain to be 
understood, e.g.\ the non-hermitian-symmetric models and their cohomological
(or other) interpretation. One would also like to find, within the present
setting, an {\em a priori} explanation for the appearance  of 
the classical cohomology of the right coset $G/H$ from the adjoint gauging
of $H$ in the bosonic sector of \ks\ models (and hence for the emergence of
quantum cohomology in suitable perturbations thereof). Here we are only able
to provide such an explanation in the fermionic sector (section 5.4). 

As regards the topologically non-trivial sectors, in particular the
torsion sectors arising from the action of $Z$ on $H$, our results
lend further support to the suggestion \cite{hori} that taking these into
account should be tantamount to resolving  the fixed point (field
identification) problem \cite{hori,fixedpoint}. In particular, we find 
that torsion sectors only arise when $m$ and $n$ are not coprime, and that
only one topological sector will ever contribute to a given 
correlation function unless $k$, $m$ and $n$ all have a common factor
(sections 4.4 and 7.1), which is precisely when one encounters problems
in the algebraic GKO approach to coset models. 

As a resolution of these problems appears to be somewhat more accessible
\cite{fixedpoint} 
in the $N=2$ models than in the bosonic models considered in 
\cite{hori}, our detailed analysis of the global aspects of \ks\ models
in \cite{bhtks} and the present paper should provide a useful basis for
testing these ideas and finding a geometrical interpretation of the
procedure suggested in \cite{fixedpoint}.

Other open issues are an understanding of level-rank duality 
at the path integral level (see \cite{hslrd} for some
relevant consideratons), and questions related
to mirror symmetry (and the corresponding topological B-models) in 
(tensor products of) \ks\ models. The results presented here and in 
\cite{bhtks} are therefore only a preliminary step
towards a full understanding of the global properties of \ks\ models.

\section{The Topological \ks\ Model}

In this section we will briefly recall the relevant results from 
\cite{bhtks}. For simplicity of exposition, these will only be presented
for $H/Z$-bundles which lift to $H$-bundles. 
In subsequent sections, these will then be 
worked out explicitly for the topological \ks\ models based on complex
Grassmannians, taking into account the additional non-trivial sectors.

\subsection{Lagrangian Realization of the Topological \ks\ Model}

Let $G$ be a compact semi-simple Lie group (which we will also
throughout assume to be simply laced), and $H$ a closed subgroup
of $G$. A Lagrangian realization for the $N=1$ super-GKO construction
\cite{gko} is provided by a gauged supersymmetric WZW model 
\cite{schnitzer} with action
\bea
S_{G/H}(g,A,\p,\pb) &=& S_{G/H}(g,A) +
   \trac{i}{4\pi}\int_{\S}\p D_{\zb}\p + \pb D_{z}\pb\non
S_{G/H}(g,A) &=& -\trac{1}{8\pi}\int_{\S}
      \g d_{A}g * \g d_{A} g - i\Gamma(g,A)\non
\Gamma(g,A) &=& \trac{1}{12\pi}\int_{M:\del M=\S}(\g dg)^{3}
\non && -\trac{1}{4\pi}\int_{\S} 
A dg\g + A \g dg + A \g A g  \label{5}\;\;.
\eea
Here $g$ is a map from the two-dimensional closed surface $\S$ to
the group $G$, $A$ is a $\lh\equiv\mbox{Lie}H$ valued gauge field for the 
(anomaly free) adjoint subgroup $H$ of $G_{L}\times G_{R}$.
$\p$ and $\pb$ are Weyl fermions taking values in the 
complexification of $\lk$, the orthogonal complement
to $\lh$ in $\lg\equiv\mbox{Lie} G$,
\be
\lg=\lh\oplus\lk\;\;,\;\;\;\;\;\;\p,\pb\in\lk^{\CC}\;\;,
\ee
so that
\bea	
D_{\zb}\p&=&\dzb\p + [\azb,\p]\non
D_{z}\pb &=&\dz \pb + [\az,\pb]\;\;.\label{dz}
\eea
Also, a trace (pairing via a non-degenerate bilinear invariant form) will
always be understood in integrals of Lie algebra valued fields.

This action has the local gauge symmetry
\be
(g,A,\p,\pb) \ra 
(h^{-1}g h,h^{-1}Ah+h^{-1}dh,h^{-1}\p h,h^{-1}\pb h)\;\;.
\ee
and an $N=1$ (actually $(1,1)$) supersymmetry (whose precise form we will 
not need to know). 

We now consider (as the most interesting class of \ks\ models),
pairs of compact Lie groups 
$(G,H)$ of equal rank, with $G$ semi-simple and such that the coset space 
$G/H$ is K\"ahler. This is equivalent to the statement that 
$H$ is the centralizer of a torus of $G$, so that it is always of the form
$H=H'\times U(1)$, where $H'$ is a product of simple factors and 
possibly further $U(1)$'s. In terms of Lie algebras this
means that, with respect to a choice of Weyl chamber of $\lg^{\CC}$,
one has a direct sum decomposition
\bea
&&\lg^{\CC} = \lh^{\CC} \oplus \lk^{+} \oplus \lk^{-}\non
&&[\lk^{\pm},\lk^{\pm}]\ss \lk^{\pm}\non
&&[\lh,\lk^{\pm}]\ss\lk^{\pm}
\eea
such that
$\lk^{+}$ is spanned by the root vectors corresponding to the positive
roots $\a\in\Delta^{+}(G/H) = \Delta^{+}(G)\setminus \Delta^{+}(H)$,
and such that $\overline{\frak{k}^{+}} = \frak{k}^{-}$. (We hope that
temporarily denoting the roots by $\a$ should not give rise to any
confusion with the fermionic field $\a$ that will appear below.)
It follows, that $\frak{k}^{\pm}$ are isotropic with respect to 
the  non-degenerate
invariant form on $\frak{g}$ which we will simply denote by $\Tr$, i.e.\
such that $\Tr ab =0$ for $a,b\in\lk^{+}$ or $a,b\in\lk^{-}$, and that
the decomposition of $\lk^{\CC}$, the complexified 
tangent space of $G/H$ at the identity element, equips $G/H$ with a
K\"ahler structure. 
In this case, the action, which in terms of the fields
\be
(\a,\b) =\Pi_{\pm}\p\;\;,\;\;\;\;\;\;(\ab,\bb)=\Pi_{\mp}\pb\;\;,
\ee
($\Pi_{\pm}$ denoting the projectors onto $\lk^{\pm}$)
becomes
\be
S_{KS}(g,A,\a,\b)=S_{G/H}(g,A) +
    \trac{i}{2\pi}\int_{\S}\b D_{\zb}\a + \bb D_{z}\ab \;\;,
\label{8}
\ee
actually has an $N=2$ supersymmetry \cite{nak,ewcp,hen1} and provides
a Lagrangian realization of the $G/H$ \ksm. The action at level $k$
is $S^{k}_{KS} = kS_{KS}$.

If $\lg$ and $\lh$ are such that
\be
[\lk,\lk]\ss\lh\;\;,
\ee
which implies that the algebras $\lk^{\pm}$ are Abelian, 
\bea
&& [\lk^{\pm},\lk^{\pm}]=0\non
&& [\lk^{+},\lk^{-}]\subset \lh\;\;,
\eea
then $G/H$ is what is known as a hermitian-symmetric space. 
In this case, $H$ is of the form $H'\times U(1)$ with $H'$ 
containing no further $U(1)$-factors.  It is
\ks\ models based on these spaces which are in a sense the easiest 
to understand and which have received the most 
attention in the literature (see e.g.\ \cite{lvw,flmw,lw}). Among
these are the complex Grassmannians which we will study in detail
in later sections.

The action of the topological \ks\ model is now obtained from the above
by what is commonly referred to as `twisting'. In the present case this
is tantamount to
regarding the fields $\a$ and $\ab$ as Grassmann odd scalars, and
$\b=\b_{z}$ and $\bb=\bb_{\zb}$ as anti-commuting $(1,0)$ and $(0,1)$
forms respectively,
\be
S_{TKS}(g,A,\a,\b)=S_{G/H}(g,A) +
    \trac{1}{2\pi}\int_{\S}\b_{z} D_{\zb}\a + \bb_{\zb} D_{z}\ab \;\;,
\label{tks}
\ee
(here we have absorbed a factor of $i$ into $\a$ and $\ab$).

As a remnant of the $N=2$ symmetry, the topological action has a
BRST-like scalar supersymmetry $\d$ (this is actually the sum
$\d=Q+\bar{Q}$ of two nilpotent supersymmetries $Q$ and $\bar{Q}$, 
which are separate invariances of the action), namely
\bea
\d g &=& g\a +\ab g\non
\d\a &=& - \trac{1}{2}[\a,\a]\non
\d\ab    &=& \trac{1}{2}[\ab,\ab]                    \non
\d\b_{z} &=& \pim(\g D_{z}g - [\a,\b_{z}])\non
\d\bb_{\zb}    &=&  \pip(D_{\zb}g \g + [\ab,\bb_{\zb}])  
\eea
Here $[.,.]$ denotes a graded commutator so that e.g.\ $[\a,\a]=2\a^{2}$.
The terms quadratic in the Grassmann-odd fields are absent if $G/H$ is
hermitian symmetric, and this is the case we will henceforth consider.

As this is, admittedly, a rather ham-handed way of introducing the
topologically twisted model, and the topological nature of this theory
is by no means obvious, let us make the following remarks, aimed
at justifying calling this the topologically twisted \ksm. 
\begin{enumerate}
\item
The twisted action differs from the supersymmetric 
action (\ref{5}) by a term of the form (spin-connection) $\times$
$R$-current, where the $U(1)_{R}$ charges of the fields taking values 
in $\lk^{\pm}$ are $\pm 1$ respectively. This is analogous to the
relation between $N=2$ sigma model and its topological twist,
the topological sigma model \cite{ewsigma} or $A$-model \cite{ewab},
and simply amounts to shifting the gauge field by half the spin connection. 
\item
It has
been shown in \cite{ns} that at the
conformal field theory level this modification of the action
amounts to twisting the energy-momentum tensor of the theory by
the $U(1)$-current $J^{N=2}$ of the $N=2$ algebra in the way defining the 
$A$-model of any $N=2$ superconformal field theory \cite{ey}, i.e.\ 
one has
\bea
T^{TKS}_{zz} &=& T^{KS}_{zz} + \trac{1}{2}\dz J^{N=2}_{z}\non
\bar{T}^{TKS}_{\zb\zb} &=& \bar{T}^{KS}_{\zb\zb} + \trac{1}{2}\dzb 
\bar{J}^{N=2}_{\zb}\;\;.
\eea
This is not obvious {\em a priori},
as  $J^{N=2}$, unlike the $R$-current of the gauge theory, is not only
bilinear in the fermions but also contains a purely bosonic part.
\item
In the Lagrangian realization we have chosen, the energy-momentum tensor
is BRST exact modulo the equations of motion of the gauge field \cite{bhtks}.
Hence correlation functions of gauge and BRST invariant and metric
independent operators are topological provided that they do not depend on the
gauge field $A$. 
\end{enumerate}

\subsection{Localization and Diagonalization of the Topological \ks\ Model}

In \cite{bhtks} it was shown that the BRST symmetry of this theory can be
used to localize the path integral to that of a bosonic $H/H$- ($H/(H/Z)$-)
like model.
Heuristically speaking, the BRST symmetry permits one to linearize the
$G/H$-part of the bosonic action, and up to a chiral anomaly the resulting
determinant cancels against that arising from the integration over the 
non-zero-modes of the Grassmann-odd fields (taking values in 
$(\lg/\lh)^{\CC}$). 

{}For notational simplicity we assume that the gauge group is (locally) of the
form $H=H'\times U(1)$ with $H'$ simple (the generalization to several
simple factors of $H'$ or additional $U(1)$'s is immediate). We have a
corresponding decomposition of the group valued field $h$ and the gauge field
$A$ as $h=(h',\exp i\varphi)$, $A=(A',A^{0})$. We denote by
$c_{G}$ and $c_{H}=c_{H'}$ the dual Coxeter numbers of $G$ and $H$,
respectively. We also define the Weyl vector of $G/H$ to be
\be
\rho_{G/H} = \trac{1}{2}\sum_{\a\in\Delta^{+}(G/H)} \a\;\;.
\ee
As 
\be
\Tr\rho_{G/H}\a=0 \;\;\;\;\;\;\forall\; \a\in\Delta^{+}(H')\;\;,
\label{rhoh}
\ee
it can be regarded
as the generator of the $U(1)$-factor of $H$.  We therefore expand
\be
A^{0} = \frac{2}{c_{G}}\rho_{G/H}a\;\;,
\ee
where $(2/c_{G})\rho_{G/H}$ is the fundamental weight of $G$ dual to the
single simple root of $G$ which is not a simple root of $H'$, since
\be
\Tr \rho_{G/H} \a = \frac{c_{G}}{2}\;\;\;\;\;\;\forall \;
\a\in\Delta^{+}(G/H)\;\;.\label{rhogh1}
\ee
{}Finally, we denote by $F_{a}=da$ the $U(1)$-part of the curvature $F_{A}$
and by $R$ the scalar curvature of the metric implicit
in the action (\ref{tks}), normalized such that
\be
\trac{1}{2\pi}\int_{\S_{g}}R = \c(\S_{g})= 2-2g
\ee
for a genus $g$ surface $\S_{g}$. Then the result of \cite{bhtks} is that the 
topological \ksm\ at level $k$ defined by the action (\ref{tks}) 
localizes to the bosonic theory with action
\bea
S^{k}_{loc}(h,A) &=& 
(k+c_{G}-c_{H'})S_{H'/H'}(h',A') + (k+c_{G})S_{U(1)/U(1)}(\exp i\varphi,a)
\non &&+ \trac{1}{2\pi} \int\Tr \rho_{G/H}\varphi R\non
&=&
(k+c_{G}-c_{H'})S_{H'/H'}(h',A') \non &&+ 
\trac{1}{2\pi}\int\Tr\varphi ((k+c_{G})F_{a} + \rho_{G/H} R)\label{hhac}
\eea
(and the integral over the Grassmann odd zero modes, if any, still to be
done).

In \cite{btver}, it was shown that the $H'$ gauge symmetry of 
the first term $S_{H'/H'}(h',A')$ can be used to abelianize the
theory, i.e.\ to reduce it to a $T'/T'$ model (plus quantum corrections)
where $T'$ is the maximal torus of $H'$ (and hence $T=T'\times U(1)$ a
maximal torus of $H$). These quantum corrections are
of three kinds. The first one is a shift of the level of the action by
$c_{H'}$.  Denoting the $T'$ and $T$ valued fields by $t'$ and
$t=(t',\exp i\varphi)=\exp i\f$, and letting $A'$ and $A=(A',a)$ from now on
stand for the torus components of the gauge fields, one thus obtains
\bea
S^{k}_{eff}(t,A) &=& (k+c_{G})S_{T'/T'}(t',A') + 
      (k+c_{G})S_{U(1)/U(1)}(\exp i\varphi, a) 
\non &&     + \trac{1}{2\pi} \int\Tr \rho_{G/H}\varphi R\non
&=& (k+c_{G})S_{T/T}(t,A)
     + \trac{1}{2\pi} \int\Tr \rho_{G/H}\varphi R\non
&=&
\trac{1}{2\pi}\int\Tr\f((k+c_{G})F_{A}+\rho_{G/H}R)\;\;.\label{ttac}
\eea

In both (\ref{hhac}) and (\ref{ttac}) we have used the fact that, in
a $U(1)/U(1)$ model the kinetic term 
$\f d*d\f$ for the compact scalar field $\f$  can be absorbed into the 
$\f da$ term by a shift $a\ra a + *d\f$ of the $U(1)$ gauge field. 
Alternatively, it suffices to note that the $a$-integral imposes
$d\f=0$, so that the kinetic term does not contribute anyway.

The second correction is a finite-dimensional determinant arising from
the ratio of the functional determinant from the $H'/T'$ components of
the gauge field and the Jacobian (Faddeev-Popov determinant) from the 
change of variables (choice of gauge) $h'\ra t'$ involved in going
from (\ref{hhac}) to (\ref{ttac}). This is the Weyl determinant
$\Delta_{W}^{H'}(t')$ appearing in the finite-dimensional 
Weyl integral formula relating integrals of class functions on $H'$ to
integrals over $T'$. It arises as a term $\sim\int R
\log\Delta_{W}^{H'}(t')$, but because of topological invariance we may
treat the fields $t'$ as position independent. Hence we can say that the
localized and abelianized theory is defined by the action (\ref{ttac})
with the modified $t'$-measure
\be
D[t']\ra D[t'](\Delta^{H'}_{W}(t'))^{\c(\S_{g})/2}\;\;.\label{weylm}
\ee

In addition there is a contribution to the action arising from the
phase of the determinant one obtains upon Abelianization which is
trivial for simply-connected gauge groups but will play a role in the
present context. It is a gauge anomaly term proportional to
$\Tr\rho_{H}F_{A}$ and, taking this into account, the total abelianized
action is
\be
S = \trac{1}{2\pi}\int\Tr\left[ \f(k+c_{G})F_{A} +\rho_{G/H}R) + 2\pi
\rho_{H}F_{A}\right]\;\;.\label{sab}
\ee
We have thus managed to reduce the original non-Abelian supersymmetric
topological \ksm\ to a much more tractable bosonic Abelian topological field
theory. 
This localized and abelianized action will be the starting point
of our investigations in the subsequent sections, the topological sectors
being accounted for by suitable constraints on $\int F_{A}$.
  
\subsection{Further Considerations}

In performing actual calculations in the model described above, 
there are some further things that are good to keep in mind, related
to chiral anomalies, fermionic zero modes and the range of integration
over the torus gauge fields $A=(A',a)$. 

Let us start by considering
the latter. The functional integral of the \ksm\ (\ref{8}) (and its
topological partner (\ref{tks})) includes by definition a sum over
all the topological sectors of $H/Z$, i.e.\ over all the isomorphism 
classes of principal $H/Z$ bundles on $\S$. If $H=H'\times U(1)$ with
$H'$ simply connected, then this means that one has to sum over all
the Chern classes of the $U(1)$ connection $a$ (with an appropriate
normalization determined by the $Z$-action on the $U(1)$-factor, and
with the obvious
generalization of this statement when there are several $U(1)$ factors
as e.g.\ in the $G/T$ models). However, even if $H'$ is simply connected,
so that there are no non-trivial $H'$-bundles, there will in general be
a further summation over (torsion) topological sectors coming from the action
of $Z$ on $H'$. How this is done will be explained in detail in the
appendix and in section 3.

In addition to this summation over topological sectors, arising from the
possibility of having non-trivial $H/Z$-bundles, there is a further source
of non-trivial topological sectors (in the abelianized theory). Namely,
it turns
out \cite{btver,btdia} that global obstructions to diagonalization of
$H'$-valued maps $h'\ra t'$ translate themselves into the requirement,
that in the Abelianized theory one has to also sum over all the topological
sectors (Chern classes) of $T'$-bundles on $\S$. Thus, in the 
final action (\ref{ttac}) all the $T$-gauge fields seem to appear on a more
or less equal footing. 

Nevertheless, the gauge field(s) associated with the explicit $U(1)$
factor(s) of $H$ continue to play a special role. To see this, we go 
back to the original action (\ref{tks}) (and to $H/Z$-bundles that
lift to $H$ bundles). 
Correlation functions in this
theory are constrained by certain selection rules arising from the
chiral anomalies of the bosonic WZW and fermionic parts of the action.
These chiral anomalies are associated with the explicit $U(1)$-factor(s)
of $H$. Let us consider, for example, the chiral transformation generated
by $\rho_{G/H}$ on the left-moving fermionic fields,
\be
(\a,\b)\ra(\ex{i\theta_{f}}\a,\ex{-i\theta_{f}}\b)\;\;.
\ee
Classically, this is an invariance of the action. However, because of
fermionic zero modes, the fermionic measure is not invariant, and one
finds a chiral anomaly proportional to the number of $\a$ minus the number
of $\b$ zero modes. By Riemann-Roch, this is determined by $\c(\S)$ and
the Chern class $c_{1}(a)$ of the $U(1)$-bundle. 
This chiral anomaly
has to be compensated for by introducing operators into the correlation
functions soaking up these zero modes.

Likewise, one can consider a chiral transformation on the bosonic part of 
the action. In terms of the variables $\f$ of the action (\ref{ttac}), this 
amounts to the chiral shift
\be
\f\ra\f + 2\theta_{b}\rho_{G/H}\;\;,
\ee
under which the action transforms as
\bea
S_{eff}^{k}(t,A)&\ra& S_{eff}^{k}(t,A) + \theta_{b}\Delta S^{k}_{eff}\non 
\Delta S^{k}_{eff} &=& 2\Tr (\rho_{G/H})^{2}
\left[\frac{2(k+c_{G})}{c_{G}}c_{1}(a) + \c(\S)\right]\;\;.\non
&=&\frac{4k}{c_{G}}\Tr (\rho_{G/H})^{2}c_{1}(a)  + 2\Tr (\rho_{G/H})^{2}
[\c(\S)+2c_{1}(a)]\;\;.
\label{deltas}
\eea
By the Freudenthal - de Vries formula and (\ref{rhoh}) one has
\be
12 \Tr (\rho_{G/H})^{2} = c_{G}d_{G}-c_{H'}d_{H'}\;\;,
\ee
where $d_{G}$ denotes the dimension of $G$. Using this identity and the
fact that for hermitian symmetric spaces one has \cite{dg2}
\be
c_{G}d_{G}-c_{H'}d_{H'} = 3c_{G}\mbox{dim}_{\CC}(G/H) \;\;,
\ee
one can rewrite (\ref{deltas}) as
\be
\Delta S^{k}_{eff} = -\trac{1}{2}k\c(\S)\mbox{dim}_{\CC}(G/H)+
(k+c_{G})\mbox{dim}_{\CC}(G/H)[\trac{1}{2}\c(\S) + c_{1}(a)]
\label{deltas2}
\ee
(see (\ref{deltas0}) for the general expression for Grassmannian $H/Z$
bundles).
The first term represents the constant background charge of the (twisted)
\ksm s, due to the anomalous transformation behavious of the $N=2$ $U(1)$
current $J^{N=2}$ in the twisted theory. 
The second term is a shift of this background charge
due to fermionic zero modes.
 
Once $c_{1}(a)$ has been fixed in terms of $\c(\S)$ by the number of 
fermionic zero modes in the correlation functions, this gives a 
$c_{1}(a)$-independent chiral anomaly which has to be compensated by
the chiral charges (weights) of the $g$-dependent observables of the
theory. 

Consider for example the sector $c_{1}(a)=-\c(\S)/2$. In that case,
there are generically no fermionic zero modes, and the chiral anomaly
of the bosonic part of the action is just
\be
\Delta S^{k}_{eff} = -\trac{1}{2}k\c(\S)\mbox{dim}_{\CC}(G/H)
\ee
in that case. This means that genus zero level one correlation functions
can possibly be interpreted as integrals over $G/H$ provided that one 
identifies an observable with chiral charge $q$ (in units of $(k+c_{G})$)
with a $2q$-form on $G/H$. This suggests the possibility to understand 
correlation functions in these models (and those related to them via level-rank
duality) in terms of the cohomology of $G/H$, as has indeed been 
argued in \cite{lvw,dg2}. This also (strongly) suggests that
such an interpretation is harder to come by in the non-hermitian symmetric
case.

\section{Global Aspects of Grassmannian Kazama-Suzuki Models}

The complex Grassmannian manifold $G(m,m+n)$ of complex $m$-planes in
$\CC^{m+n}$ can be described as the hermitian symmetric coset space 
\bea
G(m,m+n) &=& U(m+n)/U(m)\times U(n) \non
&\approx& \sumn/\sumun\;\;,
\eea
so that, in the notation of the previous section, $G=SU(m+n)$ and $H=
\sumun$.
As mentioned above, the true gauge group of this model is, as one is
gauging the adjoint action and the fermions also live in the adjoint,
the group $H/Z$ where $Z=Z(G)\cap H = \Zmn$. Topological sectors
(isomorphism classes of principal bundles of the gauge group) on a
two-dimensional closed surface $\S$ are classified by the fundamental
group of the gauge group. Thus, 
as a first step towards analyzing the Grassmannian \ks\
models based on this coset, one needs to understand the global properties 
of $H/Z$ and, in particular, determine a set of generators for 
$\pi_{1}(H/Z)$. Referring for the detailed analysis to the appendix,
we quote here the relevant results.

The fundamental group of $H/Z=\sumun/\Zmn$ is found to be
\be
\pi_{1}(H/Z) = \ZZ \times \ZZ_{\mn}\;\;,
\ee
where $\mn$ denotes the greatest common divisor of $m$ and $n$. 

{}For an element $x$ of the Lie algebra of $H$ we
will denote by $\ga_{x}$ the path $\ga_{x}(t)=\exp\tp xt$ in $H$. Loops
in $H/Z$ can be represented by (possibly open) paths in $H$ projecting
down to closed loops in $H/Z$. We also regard $H=\sumun$ as a subgroup of
$\sumn$ and denote by $\al$ and $\lal$ the fundamental roots and weights 
of $\sumn$.

The elements of the fundamental group of $\sumun/\Zmn$ can be
represented by paths $\ga_{x}$ in $\sumun$ with
\be
x = px_{free} + q x_{tor}\;\;\;\;\;\;\;\;(p,q) \in\ZZ\times \ZZ_{\mn}\;\;.
\ee
Here $x_{free}$, the generator of the free part $\ZZ$, is given by
\be
x_{free} = a\la_{m+n-1} + b\la_{1}\;\;,
\ee
with $a,b$ some solution to
\be
am+bn=\mn\;\;,
\ee
while the generator of the torsion part $\ZZ_{\mn}$ is
\be
x_{tor}=\frac{1}{\mn}(n\la_{m+n-1}-m\la_{1})\;\;.
\ee

Given the above results, it is now straightforward to work out what the
allowed Chern classes are in the various twisted sectors of the $G/H$
model. Knowledge of these will allow us to determine two of the ingredients
entering into the calculation of correlation functions, namely the index
for the fermionic zero modes and the phase of a determinant one obtains upon
diagonalization.

\subsection{Chern Classes}

The topological sectors (isomorphism classes of $H'=H/Z$ bundles) 
on a compact closed surface $\S$ are labelled by pairs
\be
(p,q)\in\ZZ\times\ZZ_{\mn}\;\;,
\ee
the $(p,q)$'th topological sector coresponding to a transition function
\be
(p,q) \LRa (\ga_{x_{free}})^{p}(\ga_{x_{tor}})^{q} \label{pqtrans}
\ee
along the boundary of some disc $D\subset\S$.
In this sector the Chern classes will be characterized by 
\be
\trac{1}{2\pi}\int_{\S}F_{A} = p x_{free} + q x_{tor}\;\;.\label{cmb}
\ee
After diagonalization, the transition functions are given by a product
of (\ref{pqtrans}) with single valued functions taking values in the
torus of $SU(m)\times SU(n)$. The topological information is contained
in the winding modes of these maps so that for the non-trivial torus
bundles one generates via diagonalization \cite{btdia} 
one can restrict one's attention to those of the form
$\ga_{x}$ with $x\in\Gamma^{r}(SU(m))\oplus\Gamma^{r}((SU(n))$. (Note that
the winding modes around $\a^{m}$ are already contained in $\ga_{x_{free}}$).

Thus after diagonalization one has the Chern classes
\be
\trac{1}{2\pi}\int_{\S}F_{A} = px_{free} + q x_{tor} + \sum_{l\neq m}n_{l}\al
\label{cmbab}
\ee
with $n_{l}\in\ZZ$. Rewriting $x_{free}$ and $x_{tor}$ in terms of roots,
this can be written more explicitly as
\be
\trac{1}{2\pi}\int_{\S}F_{A} =
\sum_{l}\left(n_{l}+p\frac{\mn + (a-b)(l-m)}{m+n}
+q\frac{l-m}{\mn}\right)\al
\label{intfa}
\ee
(with the understanding that $n_{m}=0$). This shows quite clearly which
torus bundles arise in a given topological sector $(p,q)$. It also
makes manifest the invariance of the parametrization of the topological
sectors under $q\ra q+c\mn$, $c\in\ZZ$.

On the other hand, in terms of the alternative parametrization (\ref{gt})
one has
\be
\trac{1}{2\pi}\int_{\S}F_{A} = r\a^{m}+ s\la_{m+n-1}\;\;.\label{cgt}
\ee
The claim is that, 
modulo the root lattice $\grm\oplus\grn$, 
(\ref{cmb}) and (\ref{cgt}) 
are equivalent. In order to compare them, it is helpful  to
split both expressions into their $SU(m)\times SU(n)\times U(1)$-parts.
Using various Lie algebra identities, one obtains 
(modulo the root lattice $\grm\oplus\grn$) 
\bea
(\ref{cmb}) &=& p\frac{\mn}{mn}\la_{m}\non
 &+& (pb-\frac{qm}{\mn})\lam_{1}
 + (pa+\frac{qn}{\mn})\lan_{n-1} \;\;.\label{dmb}
\eea  
and
\bea
(\ref{cgt})&=&\frac{r(m+n)+sm}{mn}\la_{m}\non
 &+& r\lam_{1} + (r+s)\lan_{n-1} \label{dgt}
\;\;.
\eea
This now permits an explicit comparison of the two parametrizations. In
establishing their equivalence, the crucial observation is that 
there is an $\mn$-fold degeneracy in the parametrisation of the 
$U(1)$-term of (\ref{dgt}), i.e.\ for each value of $r(m+n)+sm$ there 
are $\mn$ distinct pairs $(r,s)$ with $r\in\ZZ$, $s\in\ZZ_{m+n}$ giving 
that value. Thus, in particular,  a fixed value of the $U(1)$
Chern class (we will be working with later on in evaluating correlation 
functions, as this fixes the number of fermionic zero modes)
corresponds to a fixed value of $p$, with $q$ varying freely, 
while in terms of $(r,s)$ one is dealing with an $\mn$'s worth of
distinct values of $r$ and $s$.

\subsection{The Phase of the Determinant}

In calculating the ratio of determinants arising from diagonalization,
one obtains a phase factor $\exp i\Gamma$ \cite{btver,btlec}, 
\be
i\Gamma= \trac{1}{2\pi}\int \Tr 2\rho_{H}F_{A} \log(-1)\;\;,
\ee
where $\rho_{H}=\rho^{(m)} + \rho^{(n)}$ is the Weyl vector (half the sum 
of the positive roots) of $SU(m)\times SU(n)$. Writing 
\be
\log(-1) = i\pi + \tp s\;\;\;\;\;\;s\in\ZZ\;\;,
\ee
and using the representation (\ref{cmbab}) 
for $\int F_{A}$, this can be written
more explicitly as
\be
i\Gamma = \tp(2s+1)\Tr \rho_{H}(px_{free} + qx_{tor} + \sum_{l\neq m}
n_{l}\al)\;\;.
\ee
As the Weyl vector $\rho^{(m)}$ of $SU(m)$ can alternatively be written as
the sum of the fundamental weights of $SU(m)$,
\be
\rho^{(m)} = \sum_{i=1}^{m-1}\la_{i}^{(m)}\;\;,
\ee
(and likewise for $SU(n)$), the third term will always give an integral 
multiple of $\tp$ and hence not contribute to the phase. The contribution
from the first two terms can be readily determined using (\ref{intfa})
and
\bea
&&\Tr\rho_{H}\al = 1 \;\;\;\;\;\; l\neq m \non
&&\Tr\rho_{H}\a^{m} = 1 - \frac{2}{m+n}\;\;,
\eea
and one finds
\be
i\Gamma = \tp (2s+1) \left[\frac{p}{2}(a(n-1)+b(m-1)) + \frac{q}{2\mn}
(n(n-1)-m(m-1))\right]\;\;.
\ee
As the term in square brackets is always in $\frac{1}{2}\ZZ$, the ambiguity
in the branch of the logarithm is irrelevant and one actually has
\be 
i\Gamma =
\tp \left[\frac{p}{2}(a(n-1)+b(m-1)) + \frac{q}{2\mn}
(n(n-1)-m(m-1))\right] \;\;.
\ee
Thus the whole effect of the phase $\Gamma$ can be incorporated in the 
Abelianised action as a gauge anomaly term $\sim \Tr\rho_{H}F_{A}$,
\be
S=\trac{1}{2\pi}\int\Tr\left[\phi\left((k+c_{G})F_{A} +
\rho_{G/H}R\right) + 2\pi\rho_{H}F_{A}\right]\;\;,
\ee
as anticipated in (\ref{sab}).

\subsection{Index Theorem and Fermionic Zero Modes}

We now consider a fixed principal $H'=H/Z$ bundle $P'$ labelled by
$(p,q)\in\ZZ\times\ZZ_{\mn}$. We consider the $\dbar$-part of the
fermionic action given by $\int\beta_{z}^{-}\dbar_{A}\a^{+}$. 
The index of this fermionic system is given by 
\be
\mbox{Index}(\dbar_{A}) = \trac{1}{2}\mbox{dim}_{\CC}\frak{k}^{+}
\chi(\S) + c_{1}(V^{+})\;\;,\label{index1}
\ee
where $V^{+}$ is the vector bundle
\be
V^{+} = P'\times_{H'}\frak{k}^{+}
\ee
associated to $P'$ via the adjoint action of $H'$ on $\frak{k}^{+}$.
As the complex dimension of $\frak{k}^{+}$, arising as the trace of the 
identity matrix of $End_{\CC}\frak{k}^{+}$, is $mn$, the first term 
of (\ref{index1}) is simply $mn(1-g)$. The second term can be calculated
as the trace of the adjoint action of $\int F_{A}/2\pi$ on $\frak{k}^{+}$. As
$\int F_{A}$ in (\ref{cmb}) is in the torus, this action is diagonal. 
Since on a root
vector $E_{\a}\in\frak{k}^{+}$ corresponding to $\a\in\Delta^{+}(G/H)$ one 
has
\be
{}[x,E_{\a}] = (\Tr x\a) E_{\a}\;\;,
\ee
the first Chern class can be calculated to be
\bea
c_{1}(V^{+}) &=& \Tr \ad_{\frak{k}^{+}}(px_{free}+qx_{tor})\non
             &=& \sum_{\a\in\Delta^{+}(G/H)}\Tr \a (px_{free}+qx_{tor})\non
             &=& \Tr 2\rho_{G/H} (px_{free}+qx_{tor})\non
             &=& \Tr 2\rho_{G/H} px_{free}\non
             &=& \Tr (m+n)p \la_{m}x_{free} \non
             &=& p (m+n) \frac{\mn}{m+n} = p\mn\;\;.
\eea
This also follows readily from the parametrisation (\ref{dmb}), in terms
of which only the $\la_{m}$-term contributes. In any case the index is
\be
\mbox{Index}(\dbar_{A}) = mn(1-g) + p\mn\;\;.
\ee
As expected, the index does not depend on the torsion class $q$ but only 
on the winding number $p$. Combining this with a vanishing theorem, one
concludes that generically (in genus zero) 
there are no fermionic zero modes in the
topological sector $p = (g-1)mn/\mn$ ($p = -mn/\mn$), with $q$ arbitrary. 
Modulo the root lattice of $SU(m)\times SU(n)$, the corresponding generator
$px_{free}$ is equal to $(-\la_{m})$. In fact, it is easy to see directly
that they have the same winding number $\n$ (\ref{windn}),
\bea
\n(\frac{mn}{\mn}x_{free}) &=& \frac{mn}{\mn}\frac{\mn}{m+n}= \frac{mn}{m+n}
\non &=& \Tr\la_{m}^{2} = \n(\la_{m})\;\;.
\eea
Again, this also follows immediately from the parametrisation (\ref{dmb}).
Thus, in this sector the $U(1)$ gauge field can be
represented as $a\la_{m}$ with $\int da = -2\pi$. 

Thus there are an $\mn$'s
worth of torsion sectors contributing to scalar correlation functions
(meaning, correlation functions of the purely bosonic operators to be
discussed below).
In this sector with no zero modes the Chern classes can be written as
\bea
\trac{1}{2\pi}\int F_{A} &=& -\la_{m} + q x_{tor}\non
&=& -\la_{m} + \frac{q}{\mn}(n\la_{m+n-1}-m\la_{1})\non
&=& -\la_{m} + \frac{q}{\mn}(n\lan_{n-1}-m\lam_{1})\non\;\;.
\eea
{}Finally, in this sector only the torsion part of $\Gamma$ contributes to the
phase,             
\be
\ex{i\Gamma} = (-1)^{\frac{q}{\mn}(n(n-1)-m(m-1))}\;\;.
\ee

\section{The Topological Grassmannian Kazama-Suzuki Model}

\subsection{The Abelianized Action}

It follows from (\ref{ttac}), that the Abelianized action of the
Grassmannian topological \ks\ model (without the phase factor) is
(we now denote $S^{k}_{eff}$ simply by $S$)
\be
S = \trac{1}{2\pi}\int (k+m+n)\Tr \f F_{A} + \Tr\rho_{G/H}\phi R\;\;,
\ee
where the torus valued field $t$ is parametrized as $t=\exp i\f$, 
$\f$ a compact scalar field defined modulo $2\pi\Gamma^{r}$, the root
(and integral) lattice of $\sumn$. Thus, if we expand $\f$ in terms
of the simple roots of $\sumn$,
\be
\f=\sum_{l=1}^{m+n-1}\f_{l}\a^{l}\;\;,
\ee
the components $\f_{l}$ have period $2\pi$. In order to disentangle the
$U(1)$ from the $SU(m)\times SU(n)$ parts of the action, it will turn out
to be convenient to decompose $\f$ orthogonally (with respect to the trace
$\Tr$) as
\bea
\f &=& \f' + \frac{m+n}{mn}\la_{m}\f_{m} \non
\f'   &=& \fm + \fn \;\;,
\eea
where $\fm$ ($\fn$) has an expansion in terms of simpe roots of $SU(m)$
($SU(n)$). This will lead to an almost complete decoupling of the 
$SU(m)$ and $SU(n)$ sectors and thus greatly simplifies the analysis.

{}For example, for the path integral measure we will
need the Weyl determinants of $SU(m)$ and $SU(n)$. The
former is (including a factor $1/m!$, $m!$ the order of the Weyl group
of $SU(m)$)
\bea
\Delta^{(m)}_{W}(t) &=& 
\trac{1}{m!}\prod_{\a\in\Delta^{+}(SU(m))}
4\sin^{2}\trac{1}{2}\Tr\a\f\non
&=&
\trac{1}{m!}\prod_{\a\in\Delta^{+}(SU(m))}
4\sin^{2}\trac{1}{2}\Tr\a\fm    
\eea
and thus depends only on $\fm$ (and likewise for $SU(n)$). 
The normalization is such that
\be
\int_{U(1)^{m-1}}dt\Delta^{(m)}_{W}(t) = 1\;\;.\label{weyln}
\ee

Also, this decomposition is such that under the chiral shift
(chiral $U(1)$ transformation)
\be
\f \ra \f + 2\theta\rho_{G/H}\label{shift}\;\;,
\ee
which we will consider later on in the context of selection rules,
the field $\f'$ is inert while $\f_{m}$ transforms as
\be
\f_{m}\ra\f_{m}+mn\theta\;\;.
\label{shift2}
\ee 

By topological invariance (or by integration over the one-form modes
of the gauge field $A$ which imposes $d\f = 0$) we can assume that
$\f$ is constant. Hence we can use the condition (\ref{cmbab}) on the
Chern classes to rewrite the first part of the action as
\be
\trac{1}{2\pi}\int\Tr\f F_{A} = \Tr\f (
px_{free} + q x_{tor} + \sum_{l\neq m}n_{l}\al)\;\;.
\ee

By the same token, the scalar curvature term, to which only
$\f_{m}$ will contribute as
$\rho_{G/H}$ is proportional to $\la_{m}$ (\ref{rhogh}), 
becomes 
\bea
\trac{1}{2\pi}\int \Tr\rho_{G/H}\phi R &=& \trac{m+n}{4\pi}\int\f_{m}R\non
&=&(m+n)(1-g)\f_{m}\;\;,
\eea
so that the total action can be written as
\be
S = (k+m+n)\Tr\f(px_{free} + qx_{tor} + \sum_{l\neq m} n_{l}\al)
+ (m+n)(1-g)\f_{m}\;\;.\label{ksact0}
\ee
Under the chiral shift (\ref{shift}) this action changes as
\be
S\ra S + \left[kp\mn + (m+n)[(1-g)mn + p\mn]\right]\theta \label{deltas0}
\;\;.
\ee
In the sector with generically no zero modes ($px_{free}\sim (g-1)\la_{m}$),
one sees that the shift
$k\ra k+c_{G}$ cancels against the scalar curvature term and one obtains
\be
S = k(g-1)\f_{m} + (k+m+n) \Tr\f'(qx_{tor} + \sum_{l\neq m}n_{l}\al).
\label{ksact2}
\ee
Thus, under the chiral shift (\ref{shift}) the action transforms as
\be
S \ra S + kmn(g-1)\theta \;\;,\label{deltas3}
\ee
in agreement with the general result (\ref{deltas}).

\subsection{Bosonic Observables}

In \cite{bhtks}, we also determined the bosonic observables of the 
Grassmannian \ks\ models. These are functionals of
the group valued field $g$ which are invariant under the BRST
transformation
\be
\d g = g \a + \ab g
\ee
and $H$ gauge transformations. It follows readily from (\ref{obs5}) that
the upper left-hand $U(m)$ block $g^{(m)}$ of $g$ and the lower right-hand
$U(n)$ block $(g^{-1})^{(n)}$ of $g^{-1}$ are  $\d$-invariant,
\be
\d g^{(m)} =  \d (g^{-1})^{(n)} = 0\;\;.\label{obs10}
\ee 
As the remaining components of $g$ are paired exactly with the
fields $\a$ and $\ab$ under $\d$, there are no further BRST invariants
one can construct from $g$. 

It remains to impose gauge invariance. $H$ gauge transformations
act on $g^{(m)}$ and $(g^{-1})^{(n)}$ by conjugation with $h^{(m)}$
and $h^{(n)}$ respectively. Hence, a complete set of gauge and BRST
invariant operators can be obtained as traces of $g^{(m)}$ and 
$(g^{-1})^{(n)}$. For the cohomological
interpretation of the operators it turns out to be convenient
to consider as the basic set of operators the traces of $g^{(m)}$ and
$(g^{-1})^{(n)}$ in the exterior powers of the fundamental
representations of $U(m)$ and $U(n)$ respectively. We thus define
\bea
&&\cO_{a}(g) := \Tr_{\wedge^{a}}g^{(m)}\;\;,\;\;\;\;\;\;a=1,\ldots,m\;\;,
\label{obs12}\\
&&\bar{\cO}_{b}(g):=\Tr_{\wedge^{b}} (g^{-1})^{(n)}\;\;,\;\;\;\;\;\;
b = 1,\ldots,n\;\;.\label{obs13}
\eea
Since $\det g = 1$, there is one relation between these operators, namely
\be 
\det g^{(m)} \equiv \cO_{m} = \det (g^{-1})^{(n)}\equiv \bar{\cO}_{n}\;\;.
\label{obs14}
\ee
Altogether, one therefore has $\mbox{rk}(G) = m+n-1$ independent
basic gauge and BRST invariant operators generating the ring of observables
of the topological \ks\ model.

In the localized and abelianized theory, these observables reduce to the
elementary symmetric functions of the diagonal entries of $t$.
Thus, in terms of
\bea
t &=& \exp i\f = \mbox{diag}(t_{1},\ldots,t_{n+m}) \non
t_{1}&=&\ex{i\f_{1}}\non
t_{k}&=&\ex{i(\f_{k}-\f_{k-1})}\;\;,\;\;\;\;\;\;k=2,\ldots,m+n-1\non
t_{m+n} &=& \ex{-i\f_{m+n-1}}\;\;,
\eea
one has
\bea
&&\cO_{a}(\f) = \sum_{1\leq i_{1}<\ldots<i_{a}\leq m} t_{i_{1}}\ldots t_{i_{a}}
\;\;,\label{obs15}\\
&&\bar{\cO}_{b}(\f) = 
\sum_{m+1\leq j_{1}<\ldots<j_{b}\leq m+n} (t_{j_{1}}\ldots t_{j_{b}})^{-1}
\;\;.\label{obs16}
\eea

Let us consider some examples. In the $\cp{1}$ model there is one
and only one scalar operator, namely
\be
\cO_{1} = g_{11}\;\;.
\ee
In the localized theory this becomes 
\be
\cO_{1} = \ex{i\f_{1}}\;\;.
\ee
In the $\cp{2}$ model, there are two scalar operators, namely
\bea
&&\cO_{1} = g_{11} + g_{22} \ra \ex{i\f_{1}} + \ex{i(\f_{2}-\f_{1})}\non
&&\cO_{2} = g_{11}g_{22}-g_{12}g_{21} \ra \ex{i\f_{2}}\;\;.
\eea
And quite generally one finds that in the $G(m,m+n)$ models the
operator $\cO_{m}= \det g^{(m)}$ reduces to
\be
\cO_{m} = \bar{\cO}_{n} \ra \ex{i\f_{m}}\;\;.
\ee
As a final example, we consider the simplest Grassmannian which is
not a projective space, namely $G(2,4)$. In that case one has three
operators. In the localized and abelianized theory they are
\bea
&&\cO_{1}=\ex{i\f_{1}} + \ex{i(\f_{2}-\f_{1})} \non
&&\cO_{2}= \bcO_{2} = \ex{i\f_{2}}\non
&&\bar{\cO}_{1}= \ex{i\f_{3}}+ \ex{i(\f_{2}-\f_{3})}\;\;.\label{g24o}
\eea

We now need to determine the weight of these operators under the chiral
shift (\ref{shift}). It follows from (\ref{shift2}) and the explicit
expressions for the operators given above, that
\bea
\cO_{a}(\f+2\theta\rho_{G/H}) &=& \ex{ian\theta}\cO_{a}(\f)\non
\bar{\cO}_{b}(\f+2\theta\rho_{G/H}) &=& 
\ex{ibm\theta}\bar{\cO}_{b}(\f)\;\;,
\eea
so that the weights (chiral charges) of the operators are
\bea
w(\cO_{a}) &=& an\;\;\;\;\;\;a=1,\ldots,m\non
w(\bar{\cO}_{b}) &=& bm\;\;\;\;\;\;b=1,\ldots,n\;\;.\label{weights}
\eea
Hence, separating the $\f_{m}$-dependence of these operators from that
on $\f'$, we can write
\bea
\cO_{a}(\f) &=& \ex{i\frac{a}{m}\f_{m}} O_{a}(\fm)\non
\bar{\cO}_{b}(\f) &=& \ex{i\frac{b}{n}\f_{m}} \bar{O}_{b}(\fn)\;\;,
\label{ofact}
\eea
with $O_{m}=\bar{O}_{n}=1$.

\subsection{Chiral Selection Rules}
 
We are now in a position to discuss the selection rules for a general
(bosonic) correlator in the topological $\gmn$ \ksm\ at level $k$ which we will
denote by
\be
\langle\cO_{1}^{r_{1}}\ldots\cO_{m}^{r_{m}}
\bcO_{1}^{s_{1}}\ldots\bcO_{n}^{s_{n}}\rangle_{\gmnk}\;\;,\label{cor}
\ee
or simply by $\langle\ldots\rangle_{k}$ if there can be no confusion about
which
Grassmannian we are discussing. Here we take all the exponents
$r_{a}$ and $s_{b}$ to be positive, as the operators can take on
the value zero prior to abelianization. 
Also, as the correlation functions are, by standard arguments, 
independent of the positions of the operator insertions, there is no
need to indicate these insertion points explicitly in (\ref{cor}). 

One selection rule follows from considering the chiral shift (\ref{shift})
whose effect on the action we have determined before
(\ref{deltas2},\ref{deltas3}).
Thus, taking into account the weights (\ref{weights}) of the operators,
the condition for the correlation function (\ref{cor}) to be non-zero is
\be
\sum_{a=1}^{m}anr_{a} + \sum_{b=1}^{n}bms_{b} = (1-g)kmn\;\;.\label{grsel}
\ee
Since the left-hand side is manifestly positive, we see that this
condition can only be satisfied if $g=0$ (and possibly for the partition
function in genus one, but that one vanishes because of fermionic zero
modes), so we will henceforth work in genus $g=0$. 

{}For the $\cpm=G(m,m+1)$ models, (\ref{grsel}) reduces to 
\be
\sum_{a=1}^{m}ar_{a} = km\;\;.\label{cpmsel}
\ee
Let us make some observations here concerning this selection rule.
\begin{enumerate}
\item
The right-hand side is the complex dimension of the Grassmannian 
$G(m,m+k)$. As we will recall below, the cohomology of $G(m,m+n)$
is generated by $m$ Chern classes $c_{a}$ of form-degree $2a$. 
With the (tentative) identification $\cO_{a}\sim c_{a}$ the selection
rule (\ref{cpmsel}) can thus be regarded as the condition that the
correlator represent a top-form on $G(m,m+k)$. 
\item
We also note that (\ref{grsel}) reduces to (\ref{cpmsel}) if all the
$s_{b}=0$, i.e.\ if one considers chiral correlation functions in the 
$\gmnk$ model of only the $\cO_{a}$ and not the $\bcO_{b}$. This is
a special case of a more general relation between correlation functions
in $\gmn$ and $\cpm$ models which we will derive below.
\end{enumerate}

\subsection{$\Zm\times\Zn$ Selection Rules}

A further selection rule arises from a discrete symmetry of the
\ks\ model.
The original Kazama-Suzuki action naively has a discrete
$\ZZ_{m}\times\ZZ_{n}$-symmetry (acting in the obvious way via right
multiplication on the group element $g$). Less naively, this symmetry
could be broken by non-trivial bundles interfering with the $\ZZ_{m}
\times \ZZ_{n}$ directions. In the present case, these discrete 
transformations do not see the winding number sector $(-\la_{m})$, as
that one lives entirely in the $U(1)$-direction which does not interfere
with the $Z(SU(m))\times Z(SU(n))$-transformations. However, the torsion
sector could break this symmetry, and this is in fact what we will find
below. Even in the absence of torsion, however, these considerations lead
to a refinement of the selection rule arising from the invariance of the
measure and the action under the transformation $g\ra gz$. 

As the generator of $\ZZ_{m}$ is the fundamental weight $\lam_{m-1}$ of
$SU(m)$, one needs to consider the behaviour of the theory under 
\be
\fm \ra \fm + 2\pi\lam_{m-1}\;\;.
\ee
There are three contributions to consider: the action, the measure, and the
observables.

As $\Tr\la_{m}\lam_{m-1}=0$, $\Tr\lam_{m-1}\a^{l}\in\ZZ$ for 
$l\neq m$, it is obvious that the only possible
source of non-invariance of the action (\ref{ksact0}) is the torsion term.
Thus, under this shift the action changes by 
\be
2\pi(k+m+n)\Tr \lam_{m-1}qx_{tor} = -\frac{2\pi qk}{\mn} \bmod 2\pi\ZZ\;\;.
\ee
On the other hand, the observables transform as
\be
{\cal O}_{a} \ra \ex{\tp\frac{a}{m}}{\cal O}_{a}\;\;.\label{oplam}
\ee
The barred observables are invariant. 
{}Finally, the Weyl determinant (and hence the measure) is also invariant.
Thus, putting everything together one finds the condition 
\be
\sum_{a=1}^{m}\frac{ar_{a}}{m} - \frac{qk}{\mn} \in \ZZ\;\;,\label{refselq}
\ee
or
\be
\sum_{a=1}^{m} ar_{a} -\frac{q}{\mn}km = cm\;\;\;\;\;\;c\in\ZZ.
\ee
Completely analogously, one finds that the condition arising from
$\ZZ_{n}$-transformations is 
\be
\sum_{b=1}^{n} bs_{b} +\frac{q}{\mn}kn = dn\;\;\;\;\;\;c\in\ZZ.
\ee
Combining this with the selection rule
\be
\sum nar_{a} + \sum m b s_{b} = kmn\;\;,
\ee
one obtains the relation
\be
c+d=k\;\;.
\ee
This refinement of the selection rule has some interesting consequences. 
{}For instance, it implies that in the case when there is no torsion, 
$\mn=1$, the barred and unbarred operators decouple completely at
level $k=1$, as the only allowed possibilities are then $c=0,1$.

At this point one sees something special happening
when $k$ and $m$ and $n$ all have a common divisor. Namely, one can
ask the question if a given correlator can receive contributions from more
than one torsion sector. For that, one needs to be able to solve
(\ref{refselq}) simulatenously for two different values $q$ and $q'$. 
Equivalently, one is trying to solve
\be
\frac{q-q'}{\mn}k \in \ZZ\;\;.
\ee
If $(\mn|k)=1$, this has no solution for $q\neq q'$ in the allowed range.
In fact, if $\mn$ and $k$ are coprime, then 
\be
(q-q') k = c\mn\;\;\;\;\;\;c\in\ZZ
\ee
implies that $(q-q')$ is a multiple of $\mn$ (and hence zero). We will
see below that in many cases of interest only $q=0$ is allowed.

The situation is clearly  
more complicated when $k$ and $\mn$ are not coprime, a manifestation
of similar problems encountered in the conformal field theory approach in this
case \cite{fixedpoint}.

{}Finally, one could also consider $\ZZ_{m+n}=Z(G)$ transformations, 
but unsurprisingly it turns out that the resulting selection rules are 
automatically satisfied as a consequence of the relations obtained
from considering $\ZZ_{m}\times \ZZ_{n}$ plus the usual $U(1)$ chiral
selection rule.  

\subsection{Example: The $\cp{1}_{k}$ Model and the Cohomology of $\cp{k}$}

As a warm-up exercise, we now  discuss the (almost trivial) relation
between the $\cp{1}_{k}$ model and the cohomology of $\cp{k}$.
Of course, hardly any of the gyrations of the previous sections are 
needed for this example.

As there is only one operator $\cO_{1}$ in this model,  the bosonic 
genus zero correlation 
functions $\langle \cO_{1}^{r}\rangle_{k}$ are completely determined
by the selection rule (\ref{cpmsel}) which in this case reads  $r=k$.
Thus, with an approporiate normalization, we obtain \cite{ewcp}
\be
\langle\cO_{1}^{r}\rangle_{\cp{1}_{k}} = \d_{r,k}\;\;.
\ee
This means that the observable ring of the $\cp{1}_{k}$ model is
described by the single relation
\be
\cO_{1}^{k+1} = 0\;\;.
\ee
This is precisely the characterization of the
classical cohomology ring of $\cp{k}$, with the identification of
$\cO_{1}$ with the first Chern class $c_{1}$ 
of the (dual of the) tautological line bundle on $\cp{k}$
(see section 3.1). In this case, it is also readily seen that 
perturbing the action by the two-form descendant of $\cO_{1}$
\cite{ewcp,bhtks} deforms the classical cohomology of $\cp{k}$ 
to its quantum cohomology ring with defining 
relation $c_{1}^{k+1}=q$ (with $q$
the deformation parameter). This and its counterpart for other 
Grassmannians will be discussed in more detail elsewhere.

\section{Cohomology and Chiral Rings I: $\a$ Zero Modes}

In this section we will establish the first, and very direct, relationship
between the observable ring of the topological Grassmannian \ks\ models
and the cohomology ring of the corresponding Grassmannians. We will first
recall the classical description of this cohomology ring in terms of the 
Chern classes of the tautological vector bundles. We show that one
can construct operators from the Grassmann odd scalars $\a$ and $\ab$
which are BRST and gauge invariant (hence observables) and which satisfy
the cohomology relations purely algebraically, i.e.\ even outside of
correlation functions. We then show that these relations continue to be
satisfied inside genus zero correlation functions, the only contribution
coming from the topological sector $p=q=0$ in which these $\a$-operators
decouple completely from the scalar operators discussed above. 
This result can be established directly from the original action, i.e.\
without referring to the localization and abelianization of the theory.
{}Finally, we
provide a geometrical explanation of these results by identifying the
matrix $\a\ab$ with the curvature form of the tautological vector bundle.
Although we will only spell this out in detail for the Grassmannian models,
it will be evident that 
much of what follows is valid for all the hermitian symmetric \ks\ models.

\subsection{Cohomology of Grassmannians}

We briefly recall the defining relations for the cohomology
of a complex Grassmannian $G(m,m+n)$ (see e.g.\ \cite{botu})
Given a complex rank $m$ vector bundle $E$ over some manifold $M$,  
we denote by 
\be
c(E) = 1 + c_{1}(E) + c_{2}(E) + ... + c_{m}(E) 
\ee
its total Chern class. $c(E)$ satisfies the Whitney formula
\be
    c(E\oplus F) = c(E) c(F)\;\;. 
\ee
Assume now that this vector bundle can actually be regarded as a sum of 
line bundles (splitting principle). Then the Whitney formula implies that
$c(E)$ factorizes as
\be
    c(E) = (1+x_{1})(1+x_{2})....(1+x_{m})
\ee
where the $x_{i}\in H^{2}(M)$ represent (perhaps symbolically) the first
Chern classes of the individual line bundles. Alternatively, if one 
represents the Chern classes by their Weil representatives (in terms 
of curvatures of connections), the $x_{i}$ can be regarded as the 
eigenvalues of the curvature matrix of $E$. Clearly, therefore, 
the $c_{l}(E)$ are just the $l$'th elementary symmetric functions of the 
$x_{i}$,
\be
c_{l}(E) = \sum_{1\leq i_{1}<\ldots<i_{l}\leq m} x_{i_{1}}\ldots x_{i_{l}}
\;\;.
\ee

The Grassmannian $G(m,m+n)$ is the space of all complex $m$-planes in
$\CC^{m+n}$, i.e.\ every point of $\gmn$ represents a particular
$m$-plane. Attaching the corresponding $m$-plane to each point, one
obtains a rank $m$ vector bundle $E$ over $G(n,n+k)$, the 
tautological vector bundle. Likewise, one can
associate to each point the $n$-plane orthogonal to the $m$-plane it
represents. In this way, one obtains an $n$-plane bundle $F$. The fact
that at each point $x\in G(m,m+n)$ the fibres $E_{x}$ and $F_{x}$ are
canonically related by $F_{x} = \CC^{m+n}/E_{x}$ can be rephrased 
as the statement that $E$ and $F$ fit into an exact sequence
of vector bundles
\be
O\ra E \ra G(m,m+n) \times \CC^{m+n} \ra F \ra 0\;\;.
\ee
As the bundle in the center is trivial, it follows from the Whitney
formula that the total Chern classes of $E$ and $F$ are related by
\be
c(E)c(F)=1\;\;. 
\ee
The same thing is true for the sequence of dual bundles,
\bea
&&0\ra F^{*} \ra G(m,m+n) \times \CC^{m+n} \ra E^{*} \ra 0\\
&&c(E^*)c(F^*)=1\;\;.\label{cecfone}
\eea
Clearly, Chern classes of $E^*$ give rise to cohomology classes of
$G(m,m+n)$. What is less evident (but nevertheless true) is that
the cohomology is generated by these Chern classes and that the only
relations they satisfy are those following from (\ref{cecfone})
\cite{botu}. 

To be more specific, we denote by $c_{i}$, $i=1,\ldots,m$
the Chern classes of $E^*$ and by $d_{j}$, $j=1,\ldots,n$ those of $F^*$.
Then the above relation reads more explicitly
\be
(1+c_{1} + \ldots + c_{m})(1 + d_{1} + \ldots + d_{n}) = 1\;\;.\label{grrel}
\ee
This gives rise to $m+n$ relations of form degree $2,4,\ldots 2(m+n)$
respectively. The first $n$ of these are of the form $d_{j} = \ldots$
and can hence be used to express the $d_{j}$ entirely in terms of the
Chern classes $c_{i}$ of $E^*$. The remaining $m$ relations are polynomial
relations among the $c_{m}$ themselves. These will give rise to (and are
equivalently described by) all the top-degree relations of form degree
$2mn$ one can obtain from them. There will be one less top-degree relation
than the number of top-form monomials one can build from the $c_{i}$
(meaning that the relations imply the obvious fact that the top-cohomology 
group is one-dimensional). The remaining overall scale is then fixed by
the normalization
\be
\int_{\gmn}c_{m}^{n} = 1\;\;. \label{norm}
\ee

E.g.\ for $\cpm = G(m,m+1)$ one has
\be
(1+c_{1}+\ldots+c_{m})(1+d_{1})=1\;\;,
\ee
implying first of all that $d_{1}=-c_{1}$ and then that $c_{l}=c_{1}^{l}$.
Thus in this case, all the $c_{i}$ with $i\geq 2$ can also be eliminated
directly, leaving one with the one (obvious) relation $c_{1}^{m+1}=0$ for
$c_{1}$. This description of the cohomology is suitable for comparison with
the $\cpm_{1}$ \ks\ models, in which {\em a priori} we have $m$ operators
$\cO_{i}$ corresponding to the Chern classes $c_{i}$. For comparison with
the $\cp{1}_{k}$ models, on the other hand, it is more convenient (but
completely equivalent, of course) to describe the cohomology of
$\cp{k}$ using
\be
(1+c_{1})(1+d_{1}+\ldots +d_{k})=1\;\;,
\ee
leading to the elimination of the $d_{j}$ by $d_{j}=(-1)^{j}c_{1}^{j}$
and the remaining relation $c_{1}^{k+1}=0$.

\subsection{$\a$-Observables}

We will now show that from the Grassmann odd scalars $\a$ and $\ab$ we
can construct observables $C_{l}$ and $D_{l}$ satisfying the relation
(\ref{grrel}). First of all, we recall that in the hermitian symmetric models
$\ap$ and $\abm$ are BRST invariant,
\be
\d\a = \d\ab = 0\;\;,
\ee
and that they have components $\a_{ij}$ and $\ab_{ji}$ with $1\leq i\leq m$
and $m+1\leq j \leq m+n$. We can thus consider the $(m\times m)$ and 
$(n\times n)$ matrices 
\be
C=\a\ab\;\;,\;\;\;\;\;\;D=\ab\a\;\;.
\ee
Under $H$ gauge
transformations, these matrices transform in the adjoint representation
so that e.g.\ the operators
\bea
C_{a} &=& \Tr_{\wedge^{a}}C \;\;,\;\;\;\;\;\;1\leq a \leq m\non
D_{b} &=& \Tr_{\wedge^{b}}D \;\;,\;\;\;\;\;\;1\leq b \leq n
\eea
(cf.\ (\ref{obs13}))
are BRST and gauge invariant and hence qualify as observables of the 
topological Grassmannian \ks\ model. We will now show that these
operators satisfy (\ref{grrel}), i.e.\ that one has
\be
(1+C_{1}+\ldots +C_{m})(1+D_{1}+\ldots +D_{n}) = 1\;\;.\label{arel}
\ee
To that end we recall that, for any matrix $M$, one has
\be
\det (1+ M) = \sum_{k} \Tr_{\wedge^{k}}M\;\;.
\ee
Moreover, the block-diagonal matrix $\mbox{diag}(C,D)$ can be written as
the (graded) commutator of $\a$ and $\ab$,
\be
\mbox{diag}(C,D) = [\a,\ab]\;\;.
\ee
Hence, since $\Tr$ in any representation vanishes on commutators, one has
\bea
1 &=& \sum_{p=0}^{m+n}\Tr_{\wedge^{p}}[\a,\ab] \non
  &=& \det (1 + [\a,\ab]) = \det(1+\mbox{diag}(C,D)) \non
  &=& \det (1 +C) \det(1+D) \non
  &=& \sum_{a=0}^{m}\Tr_{\wedge^{a}}C \sum_{b=0}^{n}\Tr_{\wedge^{b}}D\non
  &=& (1+C_{1} + \ldots +C_{m})(1+D_{1}+\ldots +D_{n})\;\;.
\eea
We thus see that in the Grassmannian \ks\ models one can construct
operators which satisfy the cohomology relations of the corresponding
Grassmannian purely classically and algebraically, highlighting the
intimate relationship between cohomology and \ks\ models.

\subsection{Correlation Functions and Cohomology}

We will now investigate to which extent and under which conditions
the relation (\ref{arel}) continues to hold at the level of 
correlation functions. Let us therefore consider a general correlator
involving the $\a$-operators $C_{a}$ and $D_{b}$ as well as the
scalar operators $\cO_{a}$ and $\bcO_{b}$,
\be 
\langle
\prod_{a=1}^{m}C_{a}^{\gg_{a}}
\prod_{b=1}^{n}D_{b}^{\d_{b}}
\prod_{a=1}^{m}\cO_{a}^{r_{a}}
\prod_{b=1}^{n}\bcO_{b}^{s_{b}}
\rangle_{\gmn_{k}}\;\;.\label{acor}
\ee

This correlation function is constrained by conditions (selection rules)
arising from the chiral anomalies of the bosonic and fermionic parts of the
action respectively. As the operators $C_{a}$ and $D_{b}$ do not depend
on the group-valued field $g$, the former simply reads (we are now using the
unlocalized form of the action - its transformation under the chiral shift
is obtained from (\ref{deltas0}) by dropping the second term, the quantum
correction)
\be
kp\mn  + \sum_{a=1}^{m}anr_{a} + \sum_{b=1}^{n} bms_{b} = 0\;\;.
\label{asel1}
\ee
This implies that $p$ is non-positive.
In the sector $p\mn = mn(g-1)$ in which there are generically no fermionic
zero modes this selection rules reduces to that of the purely bosonic 
correlation functions discussed above, namely (\ref{grsel}). Here, on
the other hand, we are interested in the situation in which there are
$\a$ zero modes (and hence generically no $\b$ zero modes).

This drastically changes the implications of (\ref{asel1}) in the 
present context. Namely, since one wants $\a$ zero modes, i.e.\ the
index $mn(1-g) + p\mn$ to be positive, one is led to consider the
case $p=g=0$ with $mn$ zero modes, one of each species (component)
$\a_{ij}$ (there will be no $\beta$ zero modes in that case). 
We will see below that they can be interpreted directly as tangents 
to $\gmn$. 

Then the only way (\ref{asel1}) can be satisfied is when 
there are no scalar insertions whatsoever, $r_{a}=s_{b}=0$.
Thus the scalar operators $\cO_{a}$ and $\bcO_{b}$ decouple completely
from correlation functions involving $\a$ zero modes in this sector. 
{}Furthermore, the dependence on the level $k$ has disappeared.
As the fermions are inert under the action of $\Zm\times\Zn$, the
refined selection rules (\ref{refselq})
now imply that the torsion $q=0$ as well.

Each operator $C_{a}$ ($D_{b}$) soaks up $a$ ($b$) $\a$ and $\ab$
zero modes. Hence another condition required for the non-vanishing of
(\ref{acor}) is 
\be
\sum_{a=1}^{m} a \gg_{a} + \sum_{b=1}^{n} b\d_{b} = nm\label{asel3}\;\;.
\ee
With the interpretation of $C_{a}$ as a $(a,a)$-form on $\gmn$,
this is precisely the condition required to interpret the
fermionic correlator as a top-form on $\gmn$.  

It now follows that if these conditions are satisfied and if one normalizes
the correlation functions in agreement with (\ref{norm}), i.e.\ according to
\be
\langle C_{m}^{n}\rangle_{\gmn_{k}} = 1\;\;,
\ee
that the non-vanishing correlation functions are exactly the intersection
numbers of $\gmn$,
\be
\langle 
\prod_{a=1}^{m} C_{a}^{\gg_{a}}
\prod_{b=1}^{n} D_{b}^{\d_{b}}
\rangle_{\gmn_{k}} = \int_{\gmn}    
\prod_{a=1}^{m} c_{a}^{\gg_{a}}
\prod_{b=1}^{n} d_{b}^{\d_{b}}
\;\;.
\ee
Notice that this result follows directly from the classical properties of the
observables and the selection rules and that, in particular, there was no
need to invoke things like localization or diagonalization in order to 
evaluate the correlation functions.

We also want to draw attention to the fact that 
this result is independent of the level $k$. Thus for every 
$k$ there is a topological sector whose observable ring reproduces
precisely the classical cohomology ring of $\gmn$. We will find a (slightly
weaker) counterpart of this observation for bosonic correlation functions
below. Here we just want to point out that this result is rather striking
from the conformal field theory point of view, from which it is known that
the level $k=1$ chiral rings are the cohomology rings of the `target space',
while the higher level rings are rather different (unless they can be related
to $k=1$ rings via level-rank duality, as for the models based on the
projective spaces $\cpm$). It would be nice to understand this result in 
terms of Gepner's dihedrality \cite{dg2} and the spectral flow operation
considered by Nakatsu and Sugawara \cite{ns}.

\subsection{A Geometrical Interpretation} 

While we have seen above that the $\a$-operators $C_{a}$ and $D_{b}$
can be identified (already classically) with the Chern classes $c_{a}$
and $d_{b}$, this does not yet explain why such an identification is
possible. Here we will provide such an explanation by showing that
$C$ and $D$ can be regarded as the curvature forms of the tautological
vector bundles $E^{*}$ and $F^{*}$ respectively. 

Let us start in complete generality by looking
at a compact group $G$ and a compact reductive subgroup $H$. 
Thus, in terms of Lie algebras one has the decomposition 
\be
\lg = \lh \oplus \lk
\ee
with
\be
[\lh,\lk]\ss \lk\;\;.
\ee
We will first calculate the curvature of a (particular) connection
on the principal $H$-bundle $G\ra G/H$ and then specialize to pairs
$G$ and $H$ satisfying the \ks\ conditions and then further 
such that $G/H$ is hermitian symmetric.

Let $\Theta=g^{-1}\d_{G}g$ be the left-invariant Maurer-Cartan form on
$G$, and denote by $\Theta^{\lh}$ and $\Theta^{\lk}$ the components in
$\lh$ and $\lk$ respectively.  Then it is known that $\Theta^{\lh}$ is
a connection on $G\ra G/H$.  Indeed it is easy to see that this is the
case.  There are two conditions an $\lh$-valued one-form on the total
space $P$ of a principal $H$-bundle has to satisfy in order to qualify
as a  connection. The first is that evaluated on the vectors generating
the $H$-action the connection should produce the corresponding Lie
algebra element. This is true for $\Theta^{\lh}$ as the vertical part of the
Maurer-Cartan form of $P=G$. The second is that it should transform in
the adjoint representation
as one moves up and down the fibres by the principal right
action. Again this is clearly satisfied by $\Theta^{\lh}$,
$\Theta^{\lh}_{gh}= h^{-1}\Theta^{\lh}_{g} h$.

Now one can calculate the curvature of this connection using 
the Maurer-Cartan equation
\be
\d_{G}\Theta = -\trac{1}{2}[\Theta,\Theta]\;\;.
\ee
{}First of all, one has
\bea
\d_{G}\Theta^{\lh} &=& -\trac{1}{2}[\Theta,\Theta]^{\lh}\non
                 &=& -\trac{1}{2}[\Theta^{\lh},\Theta^{\lh}]
                 -\trac{1}{2}[\Theta^{\lk},\Theta^{\lk}]^{\lh}\;\;.\label{curv1}
\eea
The curvature of $\Theta^{\lh}$ is by definition the {\em horizontal}
exterior derivative of $\Theta^{\lh}$. As $\Theta^{\lh}$ is vertical and
$\Theta^{\lk}$ is horizontal, calculating the curvature is tantamount
to dropping the first term on the right hand side of (\ref{curv1}),
\bea
{}F(\Theta^{\lh}) &:=& \d^{hor}_{G} \Theta^{\lh}\non
      &=& \d_{G}\Theta^{\lh} + \trac{1}{2}[\Theta^{\lh},\Theta^{\lh}]\non
      &=& -\trac{1}{2}[\Theta^{\lk},\Theta^{\lk}]^{\lh}\;\;.\label{curv2}
\eea

Now let us make the further assumption (part of the definition of a \ks\
model) that $\lk^{\CC}$ 
decomposes into two complex conjugate subalgebras 
$\lk^{\pm}$. In this case, the curvature can be written as
\be
{}F(\Theta^{\lh}) = - [\Theta^{+},\Theta^{-}]^{\lh}\;\;.\label{curv3}
\ee
which begins to resemble the expression $[\a,\ab]$. To make this
correspondence completely explicit, consider now the case
$G=U(m+n)$ and $H= U(m)\times U(n)$.
Then $G/H = G(m,m+n)$ and what we are interested in is the curvature
of the tautological $m$-plane bundle over $G(m,m+n)$ or, what is the
same, the curvature of the principal (and tautological) 
$U(m)$-bundle over $G(m,m+n)$. Consider (\ref{curv3}) in this case.
Let the indices $i,i_{k},\ldots$ run from $1$ to $m$, and indices
$j,j_{k},\ldots$ from $m+1$ to $m+n$. Then the Maurer-Cartan forms
$\Theta^{\pm}$ have the index structure $\Theta^{+}_{ij}$ and
$\Theta^{-}_{ji}$. Thus  the $U(m)$ and $U(n)$ parts of the curvature
read 
\bea
{}F_{i_{1}i_{2}} &=& - \Theta^{+}_{i_{1}j}\Theta^{-}_{ji_{2}}\non
{}F_{j_{1}j_{2}} &=& - \Theta^{-}_{j_{1}i}\Theta^{+}_{ij_{2}}\;\;,\label{curv4}
\eea
or
\bea
&&F^{U(m)}= -\Theta^{+}\Theta^{-}\non
&&F^{U(n)}= -\Theta^{-}\Theta^{+}\;\;,\label{curv5}
\eea
with
\be
\Tr F_{i_{1}i_{2}} = -\Tr F_{j_{1}j_{2}}\label{curv6}
\ee
representing the $U(1)$-part of the curvature. 

If we now identify 
\be
\a\LRa i\Theta^{+}\;\;\;\;\;\;\ab\LRa i\Theta^{-}\;\;,
\ee
then we have the desired correspondence
\bea
&& C = \a\ab = F^{U(m)}\non
&& D = \ab\a = F^{U(n)}\;\;,
\eea
explaining why the operators $C_{a}$ and $D_{b}$ represent the Chern classes
$c_{a}$ and $d_{b}$. 

{}For completeness' sake we also mention that in the
hermitian symmetric case the BRST operator $\d$ can be identified with the
exterior covariant derivative $\d_{G}^{hor}$.
In fact, since
\bea
\d_{G} \Theta^{\pm} = - [\Theta^{\lh},\Theta^{\pm}]\;\;,\non
\d_{G} F(\Theta^{\lh}) = - [\Theta^{\lh},F(\Theta^{\lh})]\;\;,
\eea
one has
\bea
&&\d_{G}^{hor}\Theta^{\pm}=0 \LRa \d\a=\d\ab=0\;\;,\non
&&\d_{G}^{hor}F(\Theta^{\lh})=0 \LRa \d C = \d D =0\;\;.
\eea

\section{General Structure of Bosonic Correlation Functions}

We now reconsider the bosonic correlation functions (\ref{cor}). Their
structure is much richer (and more intricate) than that of their fermionic
counterparts discussed above, in particular because of the subtle level
dependence of the correlation functions. As a consequence, in this
case, we will have to work a bit harder to establish the correspondence
between these correlation functions and intersection numbers of
Grassmannians.

We thus consider a general correlation function
\be
\langle\prod_{a=1}^{m}{\cal O}_{a}^{r_{a}}\prod_{b=1}^{n}\bar{{\cal
O}}_{b}^{s_{b}}\rangle_{k,q}
\ee
in the $G(m,m+n)$ level $k$ model in the topological sector determined
by
\be
\trac{1}{2\pi}\int F_{A} = -\la_{m} + q x_{tor} + \sum_{l\neq m} n_{l}\a^{l}
\ee
(actually, we will be summing over the $n_{l}$, of course, and at some later
stage also over the allowed values of $q$, $q=0,\ldots,\mn -1$).

\subsection{The $U(1)$-Part of the Action}

As explained above, cf.\ (\ref{ksact2}),
in genus zero and in the topological sector(s) of interest the action
reduces to (treating $\f$ as a constant)
\be
S =  -k \Tr \la_{m}\f + (k+m+n)\Tr \f'(qx_{tor} + \sum_{l\neq
m} n_{l}\al)\;\;.\label{ksact}
\ee
As $\f_{m}$ appears only in the 
first term of (\ref{ksact}), to which $\f'$ does not contribute.  
and the $\f_{m}$ dependence of the observable ${\cal O}_{a}$ can be extracted
as a prefactor $\exp i \frac{a}{m}\f_{m}$ (likewise for the barred 
observables), the $\f_{m}$ integral can be done directly and its sole 
purpose in life is to impose the familiar selection rule
\be
\sum_{a=1}^{m}nar_{a} + \sum_{b=1}^{n}mbs_{b} = kmn\;\;.
\ee

{}Furthermore, the $SU(m)$ and $SU(n)$ sectors of the theory now couple 
only through the selection rule (and possibly through $x_{tor}$), and in
much of what follows we will only deal with the $\fm$-dependence explicitly,
the $\fn$-dependence being analogous.

\subsection{The $SU(m)\times SU(n)$-Part of the Action}

Let us now take a look at the third term of (\ref{ksact}), i.e.\ the term
\be
(k+m+n) \Tr\f'\sum_{l\neq m} n_{l}\al\;\;.
\ee
The summation over the $n_{l}$ can be regarded as a summation over the
root lattice $\grm\oplus\grn$ of $SU(m)\times SU(n)$. It will thus impose 
the condition that $(\kmn)\f'$ be an element of $2\pi$ times the dual lattice, 
i.e.\ the weight lattice $\gwm\oplus\gwn$,
\bea
\sum_{n_{l}}\Ra &&\trac{1}{2\pi} \fm = \frac{1}{\kmn}\lam \bmod \am\non
                &&\trac{1}{2\pi} \fn = \frac{1}{\kmn}\lan \bmod \an\;\;.
\eea
Here and in the following we use $\lam$ ($\am$) to denote a generic element 
of the weight lattice $\gwm$ (root lattice $\grm$). Thus the integral over
$\f'$ becomes a sum over a compact subset of the weight lattice of
$SU(m)\times SU(n)$ given by
\be
\trac{\kmn}{2\pi}\f'\in\left(\gwm/(\kmn)\grm\right)\times\left(\gwn/(\kmn)\grn\right)
\;\;.\label{frange}
\ee
In analogy with the observations in \cite{btver}, this implies that
correlation functions can be written in terms of sums over 
level $k+c_{G}-c_{H}$ integrable representations of $H$.

In a sense, our description of
the observables and correlation functions is quite far removed 
from the description in the conformal field theory literature
based on the cohomology of affine algebras,
see e.g.\ \cite{lvw,ht,ns}. However, it should be possible
to establish a correspondence between the two 
as the integrable level $k+c_{G}-c_{H}$ representations
of $H$ are also the fundamental
building blocks of observables in the Lie algebraic approach to
$N=2$ coset models.

\subsection{The Torsion-Part of the Action}

We now take a brief look at the torsion part of the action,
the second term of
(\ref{ksact}). Plugging $(\kmn)\f'=2\pi(\lam+\lan)$ into that term, one 
finds 
\be
(\kmn)\Tr\f'qx_{tor}=\frac{2\pi q}{\mn}\Tr(n\lan\la_{m+n-1}-m\lam\la_{1})\;\;.
\ee  
Expanding 
\bea
&&\lam=\sum n^{i}\lam_{i}  \non
&&\lan=\sum n^{m+j}\lan_{j}  \;\;,
\eea
and using
\bea
\Tr\lam_{i}\la_{1}=\frac{m-i}{m}\non
\Tr\lan_{j}\la_{m+n-1} = \frac{j}{n}\;\;,
\eea
one finds that (modulo $2\pi\ZZ$) this term becomes
\be
(\kmn)\Tr\f'qx_{tor}=\frac{2\pi q}{\mn}
(\sum in_{i} + \sum jn_{m+j})\;\;.
\ee
Combining this with the phase factor, this can be written as
\be
\Gamma + (\kmn)\Tr\f'qx_{tor}=\frac{2\pi q}{\mn}
(\sum i(n_{i}-1) + \sum j(n_{m+j}-1))\;\;.
\ee
If there were no further $q$-dependence, the sum over $q$
would now impose the condition that the weight $\la$ of $SU(m)\times SU(n)$
determined by
\be
\la+ \rho_{H} = \lam +\lan
\ee
actually gives a representation of 
\be
(SU(m)\times SU(n))/\ZZ_{\mn}\;\;.
\ee
This fact, that not all a priori possible integrable weights of 
$SU(m)\times SU(n)$ do actually appear when $\mn\neq 1$, is one
of the manifestations of the fixed point problem of conformal field
theory.

\subsection{Fourier Expansion}

In order to simplify the calculations, 
one may try to reduce the $\fm$-sum over
$\gwm/(\kmn)\grm$ to a sum over $\grm/(\kmn)\grm$. The idea is to make
use of the fact that the `difference' between the root and the weight lattice
is
\be
\gwm/\grm = Z(SU(m))= \ZZ_{m}
\ee
to write an element $\lam \in \gwm$ in some (natural) way as the sum of an  
element $\am$ of $\grm$ and a rest, the latter exponentiating to a
non-trivial element of $\ZZ_{m}$. As the center of $SU(m)$ is generated
by $\lam_{m-1}$ (which can be thought of as an element of the Lie algebra 
of $\sumn$ by embedding it as $(\lam_{m-1},0\In)$), consider the
decomposition 
\be
\lam = \am_{\la} + p_{\la}\lam_{m-1}\;\;.\label{lamam}
\ee
Here $p_{\la}\in\ZZ$. First of all, every element of the form of the right
hand side of this equation is an element of $\gwm$: the second term certainly
is, and the first term is as well since every root is a weight. However, if
$p_{\la}$ is a multiple of $m$, then $p_{\la}\lam_{m-1}$ is actually an 
element of $\grm$. In fact,      
\be 
m\lam_{m-1}=\sum_{i=1}^{m-1}i\a^{i}\;\;.\label{mlam}
\ee
Thus, to avoid an overcounting of elements of $\gwm$ in (\ref{lamam}),
we impose the condition
that $p_{\la}\in[0,\ldots,m-1]$. Finally, we need to establish 
that every element of $\gwm$ can be written in the form (\ref{lamam})
for some 
$\am_{\la}$ and $p_{\la}$. By what we know so far, this decomposition will,
if it exists, then be unique. 

Thus consider a fundamental root $\lam_{k}$. This can be written in the above 
form as
\be
\lam_{k} = \sum_{i>k} (k-i)\a^{i} + (m-k)\lam_{m-1}\;\;.
\ee
One way to verify this is to check that the right hand side satisfies the
defining relations $\Tr \a^{i}\lam_{k} = \delta^{i}_{k}$ for
$i=1,\ldots,m-1$. Thus the coefficient $p_{\la}$ in this case is
simply $(m-k)$ - which lies in the desired range. When using this result
to obtain a decomposition of this type for a general element $\lam\in\gwm$,
care needs to be taken to take the resulting $p_{\la}$, which at first
takes the value $\sum n^{k}(m-k)$ for $\lam = \sum n^{k}\lam_{k}$, 
back into the range $[0,\ldots,m-1]$ using (\ref{mlam}). But this can be
done in a unique way, and this establishes the existence and uniqueness of
the decomposition (\ref{lamam}).              

Now, the $SU(m)$-part of the partition function involves a sum over the 
elements
\bea
\trac{1}{2\pi}\fm &=& \frac{1}{\kmn}\lam \bmod \grm \non
    &=& \frac{1}{\kmn}(\am + p_{\la}\lam_{m-1}) \bmod \grm \non
    &=& \frac{1}{\kmn}\am \bmod \grm + \frac{1}{\kmn}p_{\la}\lam_{m-1}\non 
    &\in& (\ZZ_{k+m+n})^{m-1} \times \ZZ_{m}  \;\;\;\;\mbox{(as a set)}\;\;.
\label{fmaml}
\eea
Thus, by ordering this sum in such a way that for each fixed $\am\in\grm$ the
$\lam$ with $p_{\la}=0,\ldots,m-1$ appear first, it should be possible to
reduce the sum to one over the first summand by first performing the sum over 
the $p_{\la}$. 

It turns out to be simpler, however, to proceed in the opposite order.
Instead of trying to reduce the sum over $\gwm/(\kmn)\grn$ to a sum over
$\grm/(\kmn)\grm$ by summing over the $p_{\la}$, one may alternatively
try to sum over the latter for each fixed value of $p_{\la}$ and perform
the sum over the $p_{\la}$ at the end. 

Using (\ref{mlam}) and (\ref{fmaml}) 
one deduces that, in terms of the expansion 
\be
\fm = \sum_{i=1}^{m-1}\fm_{i}\a^{i} \equiv \sum_{i=1}^{m-1}N_{i}\a^{i}
\;\;,
\ee
the allowed values of $\fm$ can be parametrised as
\be
N_{i} = \frac{2\pi}{\kmn}(n_{i} + \frac{i}{m}p_{\la})\;\;,
\ee
where the $n_{i}$ range from $0$ to $\kmn -1$. 

Next, denote by $F(\fm)$ the expression
\be
{}F(\fm)= \Delta_{W}^{(m)}(\fm)\prod_{a=1}^{m-1}O_{a}^{r_{a}}(\fm)
\ee
entering in the $(m)$-sector of the correlation function. The crucial
property of $F$, following from the compactness of $\fm$, is its
periodicity,
\be
{}F(\fm) = F(\fm + 2\pi \am)\;\;\;\;\;\;\forall\;\am\in\grm\;\;.
\ee
This implies that we can expand $F$ in a finite Fourier series
(finite, because all the ingredients of $F$ are polynomials),
\bea
{}F(\fm)&=&\sum_{a\in\ZZ^{m-1}}f(a)\ex{i\sum a^{l}N_{l}}\non
      &=&\sum_{a\in\ZZ^{m-1}}f(a)\ex{\frac{\tp}{\kmn}\sum a^{l}
(n_{l}+\frac{l}{m}p_{\la})}\;\;.
\eea
($a$ can be regarded as an element of $\gwm$). 

Now the sum over the (compact range of the) $n_{l}$ will imply that
only the Fourier modes with $a^{l} = (\kmn)b^{l}$ will contribute
so that, after the sum over the $n_{l}$ one is left with
\be
(\prod_{l=1}^{m-1}\sum_{n_{l}=0}^{k+m+n-1})F(\fm) =     
\sum_{b\in(\ZZ)^{m-1}}f((\kmn)b)\ex{\tp(\sum lb_{l})\frac{p_{\la}}{m}}
\;\;.\label{prodF}
\ee

If there is no torsion, then (\ref{prodF}) is the only place where $p_{\la}$
appears. In that case, 
the sum over the $p_{\la}$ implies a further constraint
on the Fourier modes that can contribute, namely 
\be
\sum_{l=1}^{m-1}lb_{l} = 0 \bmod m\;\;.\label{slbl}
\ee
{}Frequently, this just leaves the constant mode $f(0)$ as the only non-zero
contribution to the $(m)$-sector of the correlation function, but we have
not been able to determine when this will be the case in general.

At some point, one needs to combine this with the $(n)$-sector. In the 
absence of torsion, these are only coupled via the selection rules so
the same considerations as above apply to that sector.

If there are torsion sectors then, not unexpectedly, things are slightly
different.    
{}First of all, in that case $p_{\la}$ also appears in the action, as
the only surviving contribution from the $\Tr\fm qx_{tor}$-term, the
precise form being
\[-\frac{2\pi q}{\mn}p_{\la}^{(m)}\;\;.\]
Thus the summation over the $p_{\la}$ now requires
\be
\sum_{l=1}^{m-1}lb_{l} -\frac{qm}{\mn} = 0 \bmod m
\ee
(and likewise for the $(n)$-sector).

\section{Cohomology and Chiral Rings II: Bosonic Correlation Functions}

On the basis of what we have established so far, it is now possible 
to discuss more concretely some of the general and special properties
of correlation functions in the bosonic sector. 
We will not attempt a complete and completely
explicit analysis of the correlation functions. However, we will
analyze chiral correlators and their
level-rank duality properties, and we will establish in detail the equality
of the correlation functions of the $\cp{2}_{k}$ model and the intersection
numbers of the Grassmannian manifold $G(2,k+2)$ for arbitrary $k$.

\subsection{Generalities About Chiral Correlators}

Recall that in the $\gmn$-models we have two families of bosonic operators,
the ${\cal O}_{p}$ and the $\bar{{\cal O}}_{q}$. By `chiral correlators'
we mean correlation functions depending on operators from only one of these
two families. 

The three selection rules we have found are
\bea
&& \sum_{a=1}^{m}nar_{a} + \sum_{b=1}^{n}mbs_{b} = kmn \non
&& \sum_{a=1}^{m}ar_{a} - \frac{q}{\mn} km = 0 \bmod m \non
&& \sum_{b=1}^{n}bs_{b} + \frac{q}{\mn} kn = 0 \bmod n \;\;.
\eea
Compatibility of the first with the second and third requires
that, if one has
\bea
&& \sum_{a=1}^{m}ar_{a} - \frac{q}{\mn} km = cm \non
&& \sum_{b=1}^{n}bs_{b} + \frac{q}{\mn} kn = dn \;\;,
\eea
that $c$ and $d$ be related by
\be
c+d=k\;\;.
\ee
In particular, this implies that if there is no torsion one has a decoupling
of the two chiral sectors at level $k=1$,
\be
k= \mn =1 \Ra (c=1,d=0)\mbox{   or   }(c=0,d=1)\;\;.
\ee
In general, the chiral sectors need not decouple, but for now we will just
consider chiral correlators depending only on the ${\cal O}_{a}$. For these,
the selection rules reduce to
\bea
&& \sum_{a=1}^{m}ar_{a}  = km \non
&& \sum_{a=1}^{m}ar_{a} - \frac{q}{\mn} km = cm \non
&& \frac{q}{\mn} k =  d = k-c \;\;.
\eea
Here clearly the second equality in the third equation is a consequence 
of the first two. The first equation is just the old
familiar selection rule. It coincides
with the selection rule of the $\cpm_{k}$-model (for which $\mn=1$, hence
$q=0$, so that the second and third equations are empty in that case).

The most elementary consequence of these equations is that $q=0$, $c=k$
is always a solution, so that the trivial torsion 
sector will always contribute to chiral correlators.
 
The second most elementary consequence is the condition 
that $qk/\mn$ be an integer,
\be
\frac{q}{\mn}k\in\ZZ\;\;.
\ee
This immediately implies that, if $k$ and $\mn$ are coprime, $q$ is
necessarily zero,
\be
(k|\mn)=1 \Ra q = 0\;\;.\label{kmnq}
\ee
Hence, even when $\mn\neq 1$ and, in principle, there are non-trivial
torsion sectors, these  will not contribute to chiral correlators
unless one is considering
the exceptional cases in which all three parameters $k$ and $m$ and $n$
have a non-trivial common divisor. In particular, away from these 
exceptional points (which are precisely those for which 
the field identification problems arise in conformal field theory), 
all that remains of the selection rules is 
\be
\sum_{a=1}^{m}ar_{a} = km\;\;.
\ee
If, on the other hand, $(k|\mn)\neq 1$, then non-trivial torsion sectors
can contribute as well. Clearly, the strength of the present set-up 
is that there are no major obstacles in dealing with
these models. Consider e.g.\ the case $k=m=n=2$, so that $\mn = (k|\mn)=2$
as well. Then one needs to solve
\bea
k = m = n = 2 \Ra &&r_{1} + 2r_{2} = 4\non
&& r_{1}+2r_{2} - 2q = 2c\non
&& q = 2-c\;\;.
\eea
Clearly, in this case both possible torsion sectors can contribute,
$q=0$ for $c=2$ and $q=1$ for $c=1$. 
However, as it appears to be difficult to say
anything of substance in general about the $(k|\mn)\neq 1$ correlators,
we will henceforth focus on the case $(k|\mn)=1$.

\subsection{Chiral Correlators and Level-Rank Duality for $(k|\mn)=1$}

We now consider a general chiral correlation function in the $\gmn_{k}$
model when $k$ and $\mn$ are co-prime (and hence only $q=0$ contributes). 
It follows from the (preliminary) results (\ref{prodF},\ref{slbl},\ref{kmnq})
of the previous sections that
\be
\langle\prod_{a=1}^{m}{\cal O}_{a}^{r_{a}}\rangle_{\gmn_{k}} = 
\sum_{b\in(\ZZ)^{m-1}:\sum lb_{l} =0 \bmod m} f_{m,\vec{r}}((\kmn)b)
\delta_{\sum ar_{a},km}\label{cc}
\;\;.
\ee
Here the $f_{m,\vec{r}}$ are the Fourier coefficients of 
\be
{}F(\fm)= \Delta_{W}^{(m)}(\fm)\prod_{a=1}^{m-1}O_{a}^{r_{a}}(\fm)\;\;,
\ee
so that, in particular, $\vec{r}=(r_{1},\ldots,r_{m-1})$ does not 
depend on $r_{m}$ (which only enters via the selection rules). This
will be important below.

Several immediate (and less immediate) conclusions can be drawn from 
this result:

{}First of all, $k$ and $n$ only enter in the
symmetric combination $k+m+n$ in the frequencies contributing to
(\ref{cc}), and this is the only place where $n$ appears. 
$k$, on the other hand, also appears in the Kronecker delta (selection
rule), but only in the combination $k-r_{m}$. $r_{m}$, in turn, only
appears there. 

Given this, one can read off some remarkable properties of these correlation
functions. In order not to change the sum
in (\ref{cc}), we do not want to change $m$ and $\vec{r}$, 
and we do not want to change
$k+m+n$, i.e.\ we want to keep $k+n$ fixed. Let us consider first what
happens when we exchange $k$ and $n$. The sole effect of this is to 
change the selection rule. But if we simultaneously replace $r_{m}$
by $r_{m} +  n - k$, then we retain the selection rule we started off
with,
\bea
0 &=& km - \sum_{a=1}^{m-1}ar_{a} - mr_{m} \non
  &=& nm - \sum_{a=1}^{m-1}ar_{a} - m(r_{m}+n-k) \non\;\;.
\eea
Thus, simultaneous interchange of $k$ and $n$ plus an insertion of
${\cal O}_{m}^{n-k}$ does not change the correlation functions, and
we obtain the {\bf level-rank duality}
\be
\langle\prod_{a=1}^{m}{\cal O}_{a}^{r_{a}}\rangle_{\gmn_{k}} = 
\langle{\cal O}_{m}^{n-k}
\prod_{a=1}^{m}{\cal O}_{a}^{r_{a}}\rangle_{G(m,m+k)_{n}}\;\;.
\ee
{}For $k=1$, this relates the chiral ring of the $\cpm$ model to that
of the level 1 Grassmannian model which is expected to describe its
classical cohomology ring \cite{lvw,dg1,dg2}.

To understand the necessity for the insertion of the operator ${\cal O}_{m}$,
we observe the following: Let us compare the 
couplings to the scalar curvature in the $\gmn_{k}$ and $G(m,m+k)_{n}$ models. 
In the former this is
\be
\ex{i\frac{m+n}{4\pi}\int\f_{m}R}\ra\ex{i\f_{m}(m+n)}
=\cO_{m}^{(m+n)}\;\;,\label{rmn}
\ee
while in the latter this is
\be
\ex{i\frac{m+k}{4\pi}\int\f_{m}R}\ra\ex{i\f_{m}(m+k)}
=\cO_{m}^{(m+k)}\;\;,\label{rm}
\ee
If one now inserts the operator $\cO_{m}^{(n-k)}$ into the
$G(m,m+k)_{n}$ correlation functions, then one sees that 
this just converts (\ref{rm}) into (\ref{rmn}) and thus corrects for
the difference between the gravitational background charges of the 
present topological model and the untwisted model
(for which the fermion index does not involve the scalar curvature). 

This is precisely analogous to the observation made by Witten in
\cite{ewgr}, when discussing the relation between the {\em quantum} 
cohomology ring of Grassmannians 
and the Verlinde algebra, i.e.\ correlation functions of a $G/G$ model
(see the discussion leading to eq.\ (4.63) of \cite{ewgr}). 

There, the $G/G$ model in question is a $U(m)/U(m)$
model, which essentially differs from the sector of the $\gmn$ \ksm\
we have been considering only by the absence of selection rules
due to fermionic zero modes in the former. This difference is responsible
for the drastically different ring structure of the classical and 
quantum cohomology rings. 

Quantum cohomology rings have also been
studied in the context of \ks\ models \cite{flmw,lw,lerche}, 
primarily in the language of integrable perturbations of the 
Landau-Ginzburg description of \ks\ models (these are available e.g.\ 
for level one
models of hermitian symmetric spaces based on simply-laced groups).
We will explain elsewhere what the counterpart of this approach
to quantum cohomology is in our setting.

Using the same reasoning as above we can also establish
more general dualities among correlation functions. Namely, rather than
just exchanging $k$ and $n$ we can also replace $k$ by $k+t$ and $n$ by 
$n-t$ and insert ${\cal O}_{m}^{t}$ (the choice $t=n-k$ corresponding to
what we did above). As this will also not change the selection rule,
\bea
0 &=& km - \sum_{a=1}^{m-1}ar_{a} - mr_{m} \non
  &=& (k+t)m - \sum_{a=1}^{m-1}ar_{a} - m(r_{m}+t) \;\;,
\eea
one obtains the {\bf generalized level-rank duality} (which should be 
related to Gepner's dihedrality \cite{dg2})
\be
\langle\prod_{a=1}^{m}{\cal O}_{a}^{r_{a}}\rangle_{\gmn_{k}} = 
\langle{\cal O}_{m}^{t}
\prod_{a=1}^{m}{\cal O}_{a}^{r_{a}}\rangle_{G(m,m+n-t)_{k+t}}\;\;.
\label{id3}
\ee
In particular, this shows that one can always (formally) reduce 
everything to calculations at level $k=0$, as previously observed by Gepner.
The only caveat here is that $t$ should be chosen such that $k+t$,
$m$ and $n-t$ are still co-prime as otherwise the correlation functions
in the `dual' model might also receive contributions from the torsion
sectors $q\neq 0$ which we have not considered in the above. 

While this presents one with a rather satisfactory overall picture of 
the chiral rings of \ks\ models, there are a number of things that still
remain to be understood or worked out in detail, in particular the role
of the torsion sectors in models with $(k|\mn)\neq 1$. Ideally, one would
also like to establish a general result concerning the relation between
chiral rings and cohomology rings. Lacking this, we will - in the remainder
of this article - study in detail the $\cp{2}_{k}$ model to learn about
the issues at stake.

\subsection{The Cohomology Relations for $G(2,k+2)$}

We will now apply the results of the previous
sections to the $\cp{2}_{k}$ model.
The reason why we discuss this model in detail is that here we can 
prove from scratch and for all values of the level $k$ that the 
operator ring of the model is exactly the same as the cohomology ring
of the Grassmannian $G(2,k+2)$. 

We start by taking a detailed look at the cohomology of the Grassmannian
of two-planes $G(2,k+2)$.

We choose the defining relation for the cohomology of $G(2,k+2)$ to be
\be
(1 + c_{1} + c_{2})(1 + d_{1} + \ldots + d_{k}) = 1\;\;.
\ee
{}From this the $d_{m}$ can be recursively  expressed in terms of the $c_{l}$
by the relation 
\be
d_{m} = -c_{1} d_{m-1} - c_{2} d_{m-2} \;\;.
\ee
Writing this out explicitly for some low values of $m$, one can guess a
formula for $d_{m}(c_{1},c_{2})$ which can then be proved by induction.
This formula is
\be
d_{m} = (-1)^{m}\sum_{l=0}^{[m/2]}(-1)^{l}{m-l \choose l}c_{1}^{m-2l}c_{2}^{l}
\ee
The induction relies on the identity
\be
{m-l \choose l} + {m-l \choose l-1} = {m+1-l \choose l} \;\;.
\ee
Thus the non-trivial relations among the generators $c_{1}$ and $c_{2}$ are
\bea
&& d_{k+1} = - c_{1}d_{k}  - c_{2}d_{k-1} = 0 \\
&& d_{k+2} = - c_{2} d_{k} = 0\;\;.
\eea
As the dimension of $G(2,k+2)$ is $4k$, this means that the top-degree 
relations are
\bea
I(k,m) &\equiv& c_{1}^{k-1-2m}c_{2}^{m}d_{k+1} \non
&=& \sum_{l=0}^{[(k+1)/2]}c_{1}^{2k-2(l+m)} c_{2}^{l+m} {k+1-l \choose l}
(-1)^{l} = 0\label{rel1}\\
&& m=0,\ldots, [(k-1)/2]\non
II(k,m) &\equiv& c_{1}^{k-2-2m}c_{2}^{m}(c_{2}d_{k})\non
&=&\sum_{l=0}^{[k/2]}c_{1}^{2k-2-2(l+m)} c_{2}^{l+m+1} {k-l \choose l}
(-1)^{l} = 0\;\;.\label{rel2}\\
&&m=0,\ldots, [k/2]-1 \nonumber
\eea
Thus all in all there are $k$ relations, the right number, as there are
$(k+1)$ possible top-form monomials, namely 
\be
c_{1}^{2k},c_{1}^{2k-2}c_{2},c_{1}^{2k-4}c_{2}^{2},\ldots,c_{2}^{k}\;\;.
\ee
The relations plus the normalization condition (\ref{norm}) thus determine
the integrals (intersection numbers)
\be
\int_{G(2,k+2)}c_{1}^{r_{1}}c_{2}^{r_{2}}\;\;,
\;\;\;\;\;\;\;\;\;\;\;\;r_{1}+2r_{2}=2k\;\;.
\ee
These individual integrals are somewhat difficult to extract from the
relations (\ref{rel1},\ref{rel2}). Thus, what
we will show instead below (section 8.4) is that these coincide with the 
correlation functions
\be
\langle\cO_{1}^{r_{1}}\cO_{2}^{r_{2}}\rangle_{\cp{2}_{k}}
\ee
of the $\cp{2}_{k}$ model by demonstrating that these correlators
satisfy (\ref{rel1},\ref{rel2}).  
These relations satisfy some recursion relations which will allow
us to verify that the predictions of the $\cp{2}_{k}$ model agree with
the above relations by checking the single relation $I(k,0)$ for all $k$. 
{}First of all, we observe that this is the only relation which contains
the top-degree monomial $c_{1}^{2k}$. Moreover, all the other relations 
can be expressed as powers of $c_{2}$ times relations for lower values 
of $k$.  In fact, it can easily be checked that one has 
\bea
&& I(k,m) = -c_{2} (I(k-1,m-1) + II(k-1,m-1))\;\;\;\;\;\forall\;m>0 \\
&& II(k,m) = c_{2}I(k-1,m)\;\;\;\;\;\;\forall\;m\;\;.
\eea
Therefore, everything can be reduced to checking $I(k,0)$. 
To make this slightly more explicit, say that for $k=3$ one has 
the top-degree relations
\be
c_{1}^{6} = x c_{1}^{4}c_{2} = yc_{1}^{2}c_{2}^{2} = z c_{2}^{3}\;\;.
\ee
Then the above implies that at $k=4$ one will have
\be
c_{1}^{6}c_{2}  = x c_{1}^{4}c_{2}^{2} = yc_{1}^{2}c_{2}^{3} = z c_{2}^{4}
\;\;,
\ee
and the only top-form monomial whose coefficient is not determined is
$c_{1}^{8}$. As everything starts off at $k=1$ with the relation
$c_{1}^{2}=c_{2}$, all relations are determined by the 
one relating $c_{1}^{2k}$ to the other top-form monomials, i.e.\ by
$I(k,0)$ (or, rather, for fixed $k$, by all the $I(l,0)$ with $l\leq k$).

\subsection{The $\cp{2}_{k}$ Model}

We will now apply the result (\ref{cc}) to the correlation functions
of the $\cp{2}_{k}$ model. Let us collect the ingredients that go into
the calculation.

The observables are
\bea
\cO_{1}&=&\ex{i\f_{1}} +\ex{i\f_{2}-\f_{1}} = 
\ex{i\frac{1}{2}\f_{2}}O_{1} \non
\cO_{2}&=& \ex{i\f_{2}}\;\;,
\eea
with
\be
O_{1} = \ex{i\frac{1}{2}(2\f_{1}-\f_{2})} 
+ \ex{-i\frac{1}{2}(2\f_{1}-\f_{2})}=
\ex{iN_{1}}+\ex{-iN_{1}}\;\;.
\ee
The Weyl determinant is
\be
\Delta^{(2)}_{W}(N_{1}) = 2\sin^{2}\trac{1}{2}(2\f_{1}-\f_{2}) = 2\sin^{2}N_{1}
\;\;.
\ee
In order to calculate the correlation function 
\be
\langle \cO_{1}^{r_{1}}\cO_{2}^{r_{2}}\rangle_{\cp{2}_{k}}\label{cp2c}
\ee
(subject to the selection rule $r_{1}+2r_{2} = 2k$), we have to expand
$O_{1}^{r_{1}}\Delta^{(2)}_{W}$ in a Fourier series in $N_{1}$. 
Actually, it can be easily read off from (\ref{cc}) that we can 
reduce (\ref{cp2c}) to a correlation function with only the $\cO_{1}$ insertion
at level $k-r_{2}$. This is the counterpart of the observation made
above, that at each higher $k$ the only undetermined coefficient
in the relations is that of $c_{1}^{2k}$. 

Using
\be
\Delta^{(2)}_{W}(N_{1}) = (1-\trac{1}{2}\ex{2iN_{1}} - 
\trac{1}{2}\ex{-2iN_{1}})    
\ee
and
\be
O_{1}^{r_{1}} = \sum_{l=0}^{r_{1}}{r_{1} \choose l}\ex{i(2l-r_{1})N_{1}}\;\;,
\ee
one finds 
\be
O_{1}^{r_{1}}(N_{1})\Delta^{(2)}_{W}(N_{1})=\sum_{l=-1}^{r_{1}+1}
\left[{r_{1} \choose l}-\trac{1}{2}{r_{1}\choose l+1} -\trac{1}{2}{r_{1}    
\choose l-1}\right]\ex{i(2l-r_{1})N_{1}}\;\;.
\ee
It readily follows that in this case only the constant mode $r_{1}=2l$
contributes to the correlation function and 
thus the correlation functions of the $\cp{2}_{k}$ model are 
\be
\langle \cO_{1}^{r_{1}}\cO_{2}^{r_{2}}\rangle_{\cp{2}_{k}}
= \frac{1}{r_{1}+1}{r_{1}+1 \choose \trac{1}{2}r_{1}}\d_{r_{1}+2r_{2},2k}
\;\;.\label{cp2cor}
\ee
We see that these are already correctly normalized to
\be
\langle\cO_{2}^{k}\rangle_{\cp{2}_{k}}=1\;\;.
\ee

To check that the correlation functions (\ref{cp2cor})
agree with the intersection
numbers of $G(2,k+2)$, we verify that they satisfy the relations
(\ref{rel1},\ref{rel2}), i.e. that the following identities hold: 
\bea
\sum_{l=0}^{[(k+1)/2]}
\frac{1}{k+1 -(l+m)}{ 2k - 2(l+m)\choose k-(l+m)}
{k+1-l \choose l}
(-1)^{l} &=& 0\\
 \sum_{l=0}^{[k/2]}
\frac{1}{k-(l+m)}{ 2k - 2 - 2(l+m)\choose k-1 - (l+m)}
{k-l \choose l}
(-1)^{l} &=& 0\;\;.
\eea
We also know that it is sufficient to check $I(k,0)$. Let us do this
for even $k=2j$. The argument for odd $k$ is identical.
Thus the identity we want to prove is 
\be
\sum_{l=0}^{j} \frac{1}{2j+1 -l}
{ 4j - 2l\choose 2j -l}
{2j+1-l \choose l} (-1)^{l} = 0\;\;.
\ee
Let us rewrite this as 
\be
\sum_{l=0}^{j} \frac{1}{2j+1 -2l}
{ 4j - 2l\choose 2j -l}
{2j-l \choose l} (-1)^{l} = 0\;\;,
\ee
and regard this as the value at $1$ of the function
\be
{}F_{2j}(x) = 
\sum_{l=0}^{j} \frac{1}{2j+1 -2l}
{ 4j - 2l\choose 2j -l}
{2j-l \choose l} (-1)^{l} x^{2j-2l+1} \;\;.
\ee
Our aim is to show that $F_{2j}(1)=0$. Consider first the function
\be
G_{2j}(x) = (1-x^{2})^{2j} = \sum_{l=0}^{2j}
{2j \choose 2j-l} (-1)^{l} x^{4j-2l}\;\;.
\ee
Using
\be
\frac{1}{q!}\left(\frac{d}{dx}\right)^{q} x^{p} = {p\choose q} x^{p-q}\;\;,
\ee
one obtains 
\be
\frac{1}{(2j-1)!}\left(\frac{d}{dx}\right)^{2j-1} G_{2j}(x) = 
\sum_{l=0}^{j} {2j \choose 2j-l}{4j-2l \choose 2j-1} (-1)^{l} x^{2j-2l +1}
\;\;.
\ee
Now one can check that
\be
{4j-2l \choose 2j-1}{2j \choose 2j-l} = \frac{2j}{2j-2l +1}
{4j-2l\choose 2j-l}{2j-l\choose l}\;\;,
\ee
so that finally
\be
{}F_{2j}(x) = \frac{1}{(2j)!}\left(\frac{d}{dx}\right)^{2j-1} (1-x^{2})^{2j}\;\;.
\ee
This makes it obvious that $F_{2j}(1)=0$, as there are enough powers of 
$(1-x^{2})$. In fact, for all (even and odd) values of $k$ one has
\bea
{}F_{k}(x) &=& \frac{1}{k!}\left(\frac{d}{dx}\right)^{k-1}(1-x^{2})^{k}\non
{}F_{k}(1) &=& I(k,0) = 0\;\;.
\eea
This proves the identity $I(k,0)$ and 
hence the claim that the $\cp{2}$ level $k$ model calculates the intersection
numbers of the classical cohomology ring of $G(2,k+2)$. 

Clearly, in principle one can discuss other models along these lines (and we
have verified the expected relation of the $\cp{3}_{k}$ model to the cohomology 
of $G(3,k+3)$ for certain low values of $k$); in practice, however, this is
rather cumbersome as both the explicit description of the cohomology
relations and the evaluation of the correlation functions 
become (algebraically) more and more complex.
Hence, at this point a more conceptual approach is called for, based perhaps
on the spectral flow as discussed in \cite{ns} which would allow one to
relate the bosonic correlation functions to the (more transparent) $\a\ab$
correlators discussed in section 5. We will leave this, as well as the other
issues mentioned in the introduction, to future investigations.

\subsubsection*{Acknowledgements}
 
The research of M.B. was supported by a grant from the EC within the
framework of the HCM (Human Capital and Mobility) programme. We thank
K.S. Narain for numerous discussions and L. Alvarez-Gaum\'e for
an encouraging conversation. We are also grateful to the 
referee for urging us to include the non-trivial torsion sectors 
in our discussion.

\appendix 
\section{Global Aspects of the Gauge Group $H/Z$}

\subsection{Lie Algebra Conventions for $\sumn$ and $\sumun$}

A standard choice for 
the simple roots $\al$ and fundamental weights $\lal$, $l=1,\ldots,m+n-1$,
of $\sumn$ is
\bea
&&\al = E_{l.l}-E_{l+1,l+1}\non
&&\lal= \trac{1}{m+n}\diag ((m+n-l)\II_{l},-l\II_{m+n-l})\non
&&\Tr \a^{i}\la_{k}=\delta^{i}_{\;k}\;\;,\label{sumnrw}
\eea
where $\II_{l}$ denotes the $(l\times l)$ unit matrix.
The root lattice
\be
\Gamma^{r}=\ZZ[\a^{1},\ldots,\a^{m+n-1}]
\ee
coincides with the integral lattice. The center of $\sumn$,         
\be
Z(\sumn)=\Zmn\;\;,
\ee
is given in terms of $\Gamma^{r}$ and the weight lattice 
\be
\Gamma^{w}=\ZZ[\la_{1},\ldots,\la_{m+n-1}]
\ee
by 
\be
Z(\sumn) = \Gamma^{w}/\Gamma^{r}\;\;.
\ee
It is generated by
\be
\ex{2\pi i\la_{m+n-1}} = \ex{\frac{2\pi i}{m+n}}\II_{m+n}\equiv \zmn
\ee	
and
\be
\ex{2\pi ip\la_{m+n-1}} = \ex{2\pi i\la_{m+n-p}} = \zmn^{p}\;\;.
\ee
This implies that
\be
p\la_{m+n-1}-\la_{m+n-p}\in \Gamma^{r}\;\;,\label{proot}
\ee
as can also be checked directly. Here we introduced the notation $\zmn$
for the generator of $\Zmn\ss\sumn$. Similarly, $\zm^{p}$ and $\zn^{q}$
will denote elements of $\Zm\ss SU(m)$ and $\Zn \ss SU(n)$ respectively.

It follows directly from the expression (\ref{sumnrw}) 
for the $\al$ that the Cartan matrix has components
\be
\Tr \a^{k}\a^{l} = 2 \d_{k,l}- \d_{k,l-1} - \d_{k,l+1}\;\;.
\ee
We will also need the inverse of that matrix, namely $\Tr\la_{k}\lal$.
{}From the definitions one finds, for $k<l$,
\be
\Tr\la_{k}\lal = \frac{k(m+n-l)}{m+n}\;\;.\label{lal}
\ee
In particular, one has
\bea
\Tr\la_{i}\la_{m} &=& \frac{in}{m+n}\;\;\;\;\;\;i=1,\ldots,m-1\non
\Tr\la_{m+j}\la_{m} &=& \frac{m(n-j)}{m+n}\;\;\;\;\;\;j=1,\ldots,n-1\;\;.
\label{lala}
\eea
As an application, one can use (\ref{lal}) to establish (\ref{proot}),
by showing that the traces of the left hand side of (\ref{proot}) with
all the $\lal$ are integral. In fact, one can even show slightly more
than that, namely that 
\be
p\la_{m+n-1}-\la_{m+n-p}\in\Gamma^{r}(SU(n))\equiv \grn
\ee
for $p\leq n$, and likewise that
\be
p\la_{1}-\la_{p}\in\Gamma^{r}(SU(m))\equiv\grm
\ee
for $p\leq m$. 

{}Finally, denoting as usual by $\rho_{G}$ half the sum of the positive
weights of $G=\sumn$, one has the standard result that
\be
\rho_{G}=\sum_{l=1}^{m+n-1}\lal\;\;.\label{rhog}
\ee

We regard $H=\sumun$ as a subgroup of $G=\sumn$ by embedding it in $\sumn$ 
in block diagonal form,
\be
H\ni h = \diag(h_{1}^{(m)}\ex{i\varphi/m},h_{2}^{(n)}
\ex{-i\varphi/n})\;\;,\label{h}
\ee
with $h_{1}^{(m)}\in SU(m)$ and $h_{2}^{(n)}\in SU(n)$. There is some
redundancy in this description of $h$, meaning that $H\neq SU(m)\times
SU(n)\times U(1)$, and we will come  back to this in detail below. 

The embedding is chosen such that the
\bea
&&\a^{i}\;\;\;\;\;\;i=1,\ldots,m-1\non
&&\a^{m+j}\;\;\;\;\;\;j=1,\ldots,n-1\;\;,
\eea
are the simple roots of the $SU(m)$ and $SU(n)$ subgroups of $\sumn$
respectively. Furthermore, the $U(1)$ factor is generated by the
weight
\be
\la_{m} = \trac{1}{m+n}(n\Im,-m\In)\;\;,
\ee 
as
\be
\ex{i\frac{m+n}{mn}\la_{m}\varphi}=(\ex{i\varphi/m}\Im,\ex{-i\varphi/n}\In)\;\;.
\ee
The Weyl vector $\rho_{G/H}$ (half of the sum of the positive roots
of $G$ which are not positive roots of $H$ - alternatively the sum
over all the positive roots of $G$ which contain $\a^{m}$ as a 
summand) is proportional to $\la_{m}$,
\be
\rho_{G/H}=\trac{1}{2}(n\Im,-m\In) = \trac{n+m}{2}\la_{m}\;\;.
\label{rhogh}
\ee
A useful expression for $\la_{m}$ in terms of roots is
\be
(m+n)\la_{m}=\sum_{i=1}^{m-1} in \a^{i} + mn\a^{m} +
\sum_{j=1}^{n-1}(n-j)m\a^{m+j}\;\;.\label{lam}
\ee
One observation we will need is that 
\be
\ex{\tp\la_{m}}= \zmn^{n}\;\;,\label{zmnm}
\ee
so that the subgroup of $\Zmn$ generated by $\zmn^{n}$ acts only on
the $U(1)$-part of $\sumun$. This observation will turn out to be
crucial in the analysis of the fundamental group of $\sumun/\Zmn$.

{}Finally, in order to deal with the fermions of the model, we note that
the above choice of embedding of $H$ into $G$ leads to the Lie algebra 
decomposition 
\be
\lg^{\CC} = \lh^{\CC} \oplus \lk^{+}\oplus \lk^{-}    \;\;,
\ee
where $\lk^{+}$ corresponds to the upper right hand block.
Thus, e.g.\ the Grassmann-odd scalars $\a$ and $\ab$ have components
\be
\a = (\a_{ij})\;\;,\;\;\;\;\;\;\ab = (\ab_{ji})\;\;,\;\;\;\;\;\;
i=1,\ldots,m\;\;,\;\;j=m+1,\ldots,m+n\;\;,\label{obs5}
\ee
with $\a_{ij}$ living in the root space corresponding to the positive
root
\be
\a^{i}+\ldots +\a^{j-1} \in \Delta^{+}(G/H)\;\;.
\ee
As the roots $\a\in\Delta^{+}(G/H)$ are characterized by the fact that
they contain the summand $\a^{m}$ exactly once, one has (cf.\
(\ref{rhogh1}))
\be
\Tr\a\rho_{G/H} 
= \trac{n+m}{2}
= \trac{C_{G}}{2} 
\;\;\;\;\;\;\forall\;
\a\in\Delta^{+}(G/H)\label{rho1}\;\;.
\ee

\subsection{Global Aspects of $\sumun$}

As mentioned above, there is some redundancy in the parametrization 
of $h\in \sumun$ in (\ref{h}). This redundancy means that the relation
between $\sumun$ and $SU(m)\times SU(n) \times U(1)$ is of the form
\be
\sumun = (SU(m)\times SU(n) \times U(1))/Z(m,n)
\ee
for some discrete group $Z(m,n)$. It is the purpose of this section
to determine $Z(m,n)$. Here for the first time the degree of `coprimeness'
of $m$ and $n$ will enter, as it will again in the analysis of the
action of $\Zmn$ on $\sumun$.

Redundancy means that there are distinct triples
\be
(h_{1}^{(m)},h_{2}^{(n)},\ex{i\frac{m+n}{mn}\la_{m}\varphi})
\neq
(\tilde{h}_{1}^{(m)},\tilde{h}_{2}^{(n)},
\ex{i\frac{m+n}{mn}\la_{m}\tilde{\varphi}})
\ee
representing the same element $h \in \sumun$. As a shift in $\varphi$ can 
only be compensated by an element of the center of $SU(m)\times SU(n)$,
$\Zm\times \Zn$, we clearly have
\be
Z(m,n) \ss \Zm\times\Zn\;\;.
\ee
As we also have
\be
\hm\ex{i\varphi/m}=\hm\zm^{p}\ex{i(\varphi-2\pi p)/m}
\ee
and
\be
\hn\ex{-i\varphi/n}=\hn\zn^{-q}\ex{-i(\varphi-2\pi q)/n}\;\;,
\ee
an element $(p,q)\in\Zm\times\Zn$ (thought of additively, i.e.\ as 
$(p \bmod m, q \bmod n)$ for $p,q \in \ZZ$) will be an element
of $Z(m,n)$ if there exist integers $a=a_{p,q}$ and $b=b_{p,q}$ such that
\be
\varphi - 2\pi p + 2\pi m a = \varphi - 2\pi q - 2\pi n b\;\;,
\ee
i.e.\ such that
\be
p-q = ma + nb \label{cond1}
\ee
for some $a,b \in \ZZ$. At this point some elementary number theory
enters the game. Denote by $\mn$ the greatest common divisor 
of $m$ and $n$,
\be
\mn = \mbox{gcd}(m,n)\;\;.
\ee
The fact we will need is that for any $a,b \in \ZZ$
\be
am + bn = 0 \bmod \mn
\ee
and that for any integer $c\in \ZZ$ one can find $a_{c},b_{c}\in\ZZ$
such that
\be
a_{c}m + b_{c}n = c \mn\;\;.
\ee
This fact is usually stated in the form that for coprime integers
$m_{0}$ and $n_{0}$ one can find $a,b \in\ZZ$ such that
\be
am_{0} + b n_{0} = 1\;\;.
\ee
Thus there is a solution to (\ref{cond1}) 
for some $a,b\in\ZZ$ iff $p-q = 0 \bmod \mn$ and one has
\be
Z(m,n) = \{(p,q)\in\Zm\times\Zn:p-q=0\bmod \mn\}\;\;.
\ee
It is relatively easy to see that this can be described more
explicitly as
\be
Z(m,n) = \ZZ_{mn/\mn}\;\;.\label{zmn}
\ee
One way of doing this is to show that the element $(p,q)=(1,1)$ is a
generator of $Z(m,n)$, its order obviously being the least common
multiple of $m$ and $n$, i.e.\ $mn/\mn$.
 
The upshot of this is that the relation 
between $\sumun$ and $SU(m)\times SU(n) \times U(1)$ is
\be
\sumun = (SU(m)\times SU(n) \times U(1))/\ZZ_{mn/\mn}\;\;.
\ee
As an example consider $(m,n=1)$, i.e.\ $\cpm$. In that case,
we have $\mn=1$, $mn/\mn = m$, and one recovers the familiar result
\be
S(U(m)\times U(1)) = (SU(m) \times U(1))/\Zm = U(m)\;\;.
\ee

\subsection{The Fundamental Group of $\sumun/\Zmn$}

The relevant gauge group in the $G/H$ 
gauged Wess-Zumino-Witten models is not $H$ but rather
\be
H' = H/(Z(G)\cap H)\;\;,
\ee
(not to be confused with the group we called $H'$ in section 2)
where $Z(G)$ denotes the center of $G$, as the gauge group has to be 
a subgroup of the adjoint group $G/Z(G)$ of $G$. The aim will now be
to 
\begin{enumerate}
\item study the action of $Z(G)\cap H$ on $H$;
\item determine the fundamental group $\pi_{1}(H')$;
\item exhibit an explicit set of generating paths for $\pi_{1}(H')$.
\end{enumerate}
Note that for our purposes it is not enough to just determine
$\pi_{1}(H')$ by some means.  It is (3) that we need in order to be
able to explicitly introduce the twisted topological sectors into the
path integral via transition functions (equivalently, via conditions
on the allowed Chern classes).

{}For $G=\sumn$, $Z(G)=\Zmn$  is generated by
\be
\zmn = \ex{\frac{\tp}{m+n}}\II_{m+n} = \ex{\tp\la_{m+n-1}}\;\;.
\ee
But clearly $\zmn$ is an element of $H=\sumun\ss \sumn$, as both blocks
of $\zmn$ are unitary, and the determinant of $\zmn$ is equal to one.
Hence we have
\be
H\cap Z(G) = Z(G) = \Zmn \;\;,
\ee
and
\be
H' = \sumun/\Zmn\;\;.
\ee
Now, let us recall the observation (\ref{zmnm}) that
\be
\zmn^{n} = \ex{\tp\la_{m}}\;\;.
\ee
This means that the subgroup of $\Zmn$ generated by $\zmn^{n}$ acts only
on the $U(1)$-part of $\sumun$ which is generated by $\la_{m}$. The
reason why it is necessary to disentangle this subgroup form the rest is
that it essentially does not change the fundamental group. `Essentially'
means that $\pi_{1}(U(1))$ is isomorphic to $\pi_{1}(U(1)/\ZZ_{r})$, the
isomorphism being given by 
\bea
&&\pi_{1}(U(1))=\ZZ \ni 1 \rightarrow r \in \pi_{1}(U(1)/\ZZ_{r}) =\ZZ\non
&&\pi_{1}(U(1)/\ZZ_{r})/\pi_{1}(U(1)) = \ZZ/r\ZZ = \ZZ_{r}\;\;.\label{pisim}
\eea
Thus the question is what is the order of the group generated by $\zmn^{n}$
or, in other words, what is the smallest positive integer $p$ such that
$\zmn^{np}=1$. Now
\be
\zmn^{np}=1 \LRa \frac{pn}{m+n}\in\ZZ\;\;.
\ee
To determine the smallest integer for which this is the case, we once
again write $m= m_{0}\mn$, $n=n_{0}\mn$, and try to find the smallest
solution $p$ to
\be
pn = a (n+m) \;\;\;\;\;\;a\in\ZZ\;\;,
\ee
or
\be
pn_{0} = a (m_{0}+n_{0})\;\;.
\ee
As $n_{0}$ and $m_{0}+n_{0}$ have no common factor, the `minimal' solution
is $a=n_{0}$ and $p=m_{0}+n_{0}$ or $p=(m+n)/\mn$. Thus the subgroup of
$\Zmn$ acting only on the $U(1)$ is
\be
\{\zmn^{np},p\in\ZZ\} \approx \ZZ_{(m+n)/\mn}\;\;,\label{zmpn}
\ee
the generator
\be
\zmn^{np}=\zmn^{\mn}
\ee
of $\ZZ_{(m+n)/\mn}$ being obtained for $p=b-a$ where $am+bn=\mn$. 

In particular, in the coprime case $\mn=1$, this is all of $\Zmn$,
in line with the conformal field theory contention that in this 
case no complicating features due to fixed points and/or topological 
sectors should arise.

In general, the result (\ref{zmpn})
implies that the group giving rise to new
topological sectors in $H'$ is the quotient group
\be
\Zmn/\ZZ_{(m+n)/\mn}= \ZZ_{\mn}\;\;.
\ee
The above discussion suggests (and almost proves - we will not fill in 
the missing steps) that the fundamental group of $H'$ is
\be
\pi_{1}(\sumun/\Zmn)=\ZZ \times \ZZ_{\mn}\;\;,\label{pi1}
\ee
and we will now proceed to exhibit explicitly a set of generators.

As principal $H'$-bundles on a surface $\S$ are labelled by
$\pi_{1}(H')$, what (\ref{pi1}) shows is that 
\begin{itemize}
\item new topological 
sectors occur precisely when $m$ and $n$ are not coprime;
\item these are always torsion, so that there are a finite 
number of new topological sectors to be taken into account;
\item the sum over the new topological sectors is a finite sum over 
the elements of $\ZZ_{\mn}$. 
\end{itemize}

\subsection{Generating Loops of the Fundamental Group of $\sumun/\Zmn$}

As mentioned before, for our purposes it is not sufficient to just determine
$\pi_{1}(H')$ in some way (as above). What we need in order to implement
the topological sectors in the path integral and to read off the
corresponding Chern classes is an explicit set of representatives of the 
generators of $\pi_{1}(H')$. As we are working at the level of $H$, this
amounts to a suitable set of paths in $H$ which are closed in $H$ up to
the action of $H \cap Z(G) = \Zmn$. 

A remark on notation. For an element $x$ of the Lie algebra of $H$ we
will denote by $\ga_{x}$ the path $\ga_{x}(t)=\exp\tp xt$ in $H$. We
will also call {\em winding number \/} of $\ga_{x}$ the real number
\be
\n(\ga_{x}) \equiv \n(x) = \Tr\la_{m}x\;\;.\label{windn}
\ee
This winding number is sensitive only to the free part of the fundamental
group (living in the $U(1)$-factor) and not to the torsion part.
Of course, this is only literally a winding number if $\ga_{x}$ is
a closed path in $H$ or $H'$ (and need not be an integer in the latter case). 
Thus, single valued loops
$\ga_{x}$ with $x\in \Gamma^{r}$ give integer winding numbers, as in
\be
x=\sum_{l} n_{l}\a^{l} \Ra \n(\ga_{x})=n_{m}\;\;.
\ee

We will first determine a generator of the fundamental group of $\sumun$.
With the parametrization 
\be
h=(\hm\ex{i\f/m},\hn\ex{-i\f/n})\;\;,
\ee
we had the identification
\be
\sumun = (SU(m)\times SU(n)\times U(1))/\ZZ_{mn/\mn}\;\;.\label{sumunzmn}
\ee
This is reflected in the fact that the minimal closed loop living entirely
in the $\f$-direction,
\be
g(t) = \ex{\tp\frac{m+n}{\mn}\la_{m}t}
=(\ex{\tp\frac{n}{\mn}t}\Im,\ex{-\tp\frac{m}{\mn}t}\In)
\ee
has winding number
\be
\n(x=\frac{m+n}{\mn}\la_{m}) = \frac{m+n}{\mn}\Tr\la_{m}^{2}=\frac{mn}{\mn}
\;\;.
\ee
While this is thus the image of the generator of $\pi_{1}(U(1))$ under the
identification (\ref{sumunzmn}), this is not the generator of $\pi_{1}(H)$.
To see this, consider the loop
\be
\ga_{\a^{m}}(t)=\ex{\tp \a^{m}t}\;\;.
\ee
Clearly, this loop has winding number one,
\be
\n(\a^{m})=1\;\;,
\ee
and can be chosen to be the generator of $\pi_{1}(\sumun)$.
Moreover, using (\ref{lam}), one can verify that
\be
\frac{m+n}{\mn}\la_{m} = \frac{mn}{\mn}\a^{m} \bmod \Gamma^{r}(SU(m))\oplus
\Gamma^{r}(SU(n)) \;\;,
\ee
so that 
\be
(\ga_{\a^{m}})^{\frac{mn}{\mn}}(t) \sim g(t)\;\;,
\ee
where $\sim$ means 'homotopic to'. This holds because the terms depending
on the roots of $SU(m)$ and $SU(n)$ will give contractible loops in 
$\sumun$. What the above means in terms of bundles is that a principal
$H$ bundle constructed with $\ga_{\a^{m}}^{p}$ as a transition function will
not lift to a principal $SU(m)\times SU(n) \times U(1)$ bundle unless
$p$ is a multiple of $mn/\mn$.

We now come to the thornier issue of generators for $\pi_{1}(H')$.
{}From the point of view of $H$, these are certain paths in $H$ which
start at the identity element and end at an element of $\Zmn\ss\sumun$,
so that they define non-trivial closed loops in $H'=H/\Zmn$. 

What we expect to find are
a generator of the free part $\ZZ\ss\pi_{1}(H')$ which is such that its
$(m+n)/\mn$'th power is homotopic to the loop $\ga_{\a^{m}}$, the
generator of $\pi_{1}(H)$, and a generator of the torsion part 
$\ZZ_{\mn}\ss\pi_{1}(H')$ which is such that its $\mn$'th power is
contractible. 

{}First of all, recall that $\Zmn$ is generated by
\be
\zmn = \ex{\tp\la_{m+n-1}}
\ee
with 
\be
\zmn^{p}=\ex{\tp p\la_{m+n-1}}=\ex{\tp\la_{m+n-p}}\;\;.
\ee
It is thus natural to parametrize the loops of interest (ending at some
point of $\Zmn$) as $\ga_{x}$ with $x\in \Gamma^{w}$, the weight lattice.

We begin with the free part.  
Up to homotopy, the most general loop of the type $\ga_{x}$ with
$x\in\Gamma^{w}$ is of the form $x=a\la_{m+n-1}+b\la_{1}$
since, as noted before, any other weight is equal to a multiple of 
either $\la_{1}$ or $\la_{m+n-1}$ modulo the root lattice of 
$SU(m)\times SU(n)$. Thus, let us consider the loop $\ga_{x}$ with
\be
x = a\la_{m+n-1} + b\la_{1}\;\;.
\ee
Its winding number is
\be
\n(\ga_{x})= \frac{am+bn}{m+n}\;\;.
\ee
Thus, using again the general result that 
\be
\mbox{min}_{a,b}(am+bn) = \mn\;\;,
\ee
we see that the minimum winding number is
\be
\mbox{min}_{x}\n(\ga_{x})=\frac{\mn}{m+n}\;\;,
\ee
which is obtained for $x=x_{free}$,
\be
x_{free}=a\la_{m+n-1} + b\la_{1}\;\;,\;\;\;\;\;\;am+bn = \mn\;\;.
\ee
But, comparing with the generator of the fundamental group of $H$,
this is precisely what we expect as this means that we get an
$\frac{m+n}{\mn}$'s worth of new generators for the free part of 
$\pi_{1}(H')$, corresponding to the $\ZZ_{(m+n)/\mn}$ subgroup of
$\Zmn$ acting on the $U(1)$-part of $H$. 

A solution to $am+bn=\mn$ is only unique up to the shift $a\ra a+cn/\mn$,
$b\ra b-cm/\mn$. Under this shift $x_{free}$ transforms as
\be
x_{free}\ra x_{free} + c(\frac{n}{\mn}\la_{m+n-1}-\frac{m}{\mn}\la_{1})\;\;.
\ee
We will see below that the term in brackets represents the
generator of the torsion subgroup $\ZZ_{\mn}$ of $\pi_{1}(H')$, so the 
ambiguity in the choice of $a$ and $b$ only amounts to a mixing of the
generators. 

We now come to the torsion part.
We are still working with paths in $H$ which are of the form $\ga_{x}$ with
$x\in\Gamma^{w}$, as these correspond to non-trivial loops in $H'$. Also, 
as before, we can restrict our attention to $x$ of the
form $x = a \la_{m+n-1} + b\la_{1}$. 
A torsion loop should not be detectable by the $U(1)$ winding number
measured by $\n(\ga_{x})$, so that one is looking for non-trivial solutions
to
\be
\n(\ga_{x})=0 \LRa am + bn =0 \LRa am_{0} +bn_{0} = 0\;\;.
\ee
As $m_{0}$ and $n_{0}$ are coprime, i.e.\ have no common factors, this
can only have a solution if $a$ has a factor of $n_{0}$ and $b$ has a
factor of $m_{0}$. In fact, the general solution is
\be
(a,b) = (cn_{0},-cm_{0})\;\;\;\;\;\;\;\;c\in\ZZ\;\;.
\ee
The minimal solution $c=1$ corresponds to the element
\be
x_{tor} = n_{0}\la_{m+n-1}-m_{0}\la_{1}=
\frac{n}{\mn}\la_{m+n-1}-\frac{m}{\mn}\la_{1} \in\Gamma^{w}\;\;.
\ee
It has the crucial property that
\be
\mn x_{tor} \in\Gamma^{r}(SU(m))\oplus\Gamma^{r}(SU(n))\;\;,\label{xtort}
\ee
so that the loop $\ga_{\mn x_{tor}}$ is homotopically trivial,
\be
\ga_{\mn x_{tor}}=(\ga_{x_{tor}})^{\mn} \sim 0 \in \pi_{1}(H')\;\;,
\label{xtort2}
\ee
precisely the property one would expect of the generator of a $\ZZ_{\mn}$
subgroup of $\pi_{1}(H')$. To establish (\ref{xtort}), we recall 
that both $m\la_{1}$ and $n\la_{m+n-1}$ are equivalent to
$\la_{m}$ modulo $\Gamma^{r}$. Hence their difference is zero modulo 
$\Gamma^{r}$ and defines a trivial loop in $H$ (and hence, in particular,   
in $H'$). 

In summary, the elements of the fundamental group of $\sumun/\Zmn$ can be
written as paths $\ga_{x}$ in $\sumun$ with
\be
x = px_{free} + q x_{tor}\;\;\;\;\;\;\;\;(p,q) \in\ZZ\times \ZZ_{\mn}\;\;.
\ee

There is an alternative approach to determining the 
transition functions (and hence Chern classes) of $\sumun/\Zmn$-bundles 
which is simpler and more direct but 
makes less manifest the group structure of the fundamental group. 
One starts with $\sumun$-bundles and
the simple result that for these the generator can be
chosen to be $\a^{m}$. One then notes 
that the allowed transition functions for $\sumun/\Zmn$-bundles are
then necessarily of the form 
\be
r\a^{m}+s\la_{m+n-1}\;\;,\;\;\;\;\;\;(r,s)\in\ZZ\times \ZZ_{m+n}\label{gt}
\ee
where $\la_{m+n-1}$ is the generator of $\Zmn$ and hence $s=0,\ldots,m+n-1$.
This should be compared with the (rather more complicated) expression
\bea
&&px_{free} + qx_{tor}\;\;\;\;\;\;\;\;(p,q)\in\ZZ\times\ZZ_{\mn}\non
&&x_{free} = a\la_{m+n-1} + b\la_{1}\;\;\;\;\;\;\;\;am+bn=\mn\non
&&x_{tor}=\frac{1}{\mn}(n\la_{m+n-1}-m\la_{1})\label{mb}
\eea
obtained above. 
It can be shown with some work that these two parametrizations are
(despite appearance) equivalent and we will briefly indicate the relation
between the two in section 3.1. 

\rnc{\Large}{\normalsize}

\end{document}